\documentclass[aps,prx,reprint,superscriptaddress]{revtex4-2}

\usepackage{graphicx}
\usepackage{dcolumn}
\usepackage{bm,braket,here,comment,color}
\usepackage{mathtools,amssymb,hyperref}
\allowdisplaybreaks[1]

\renewcommand\d{\partial}
\newcommand\+{\dagger}
\newcommand\<{\langle}
\renewcommand\>{\rangle}
\renewcommand\r{{\bm{r}}}
\newcommand\s{{\bm{s}}}
\newcommand\eff{\mathrm{eff}}
\newcommand\reg{\mathrm{reg}}

\begin{document}

\preprint{}

\title{Exploring the Strongly-Interacting Regime of Effective Multi-Body Interactions\\
in a Trapped Ultracold Atom System}


\author{Kantaro Honda}
 \email[]{honda.kantaro.35m@st.kyoto-u.ac.jp}
\altaffiliation{}
 \affiliation{Department of Physics, Graduate School of Science, Kyoto University, Kyoto 606-8502, Japan}

\author{Yosuke Takasu}
  \affiliation{Department of Physics, Graduate School of Science, Kyoto University, Kyoto 606-8502, Japan}

\author{Yuki Haruna}
  \affiliation{Department of Physics, Graduate School of Science, Kyoto University, Kyoto 606-8502, Japan}

\author{Yusuke Nishida}
  \affiliation{Department of Physics, Institute of Science Tokyo, Ookayama, Meguro, Tokyo 152-8551, Japan}

\author{Yoshiro Takahashi}
  \affiliation{Department of Physics, Graduate School of Science, Kyoto University, Kyoto 606-8502, Japan}


\date{\today}

\begin{abstract}
A two-body interaction or force between quantum particles is ubiquitous in nature, and the microscopic description in terms of the bare two-body interaction is the basis for quantitatively describing interacting few- and many-body systems.
Alternatively, the effective description in terms of an effective two-body interaction successfully captures the essence of the systems.
However, for several important observations, the explanation in terms of an effective two-body interaction is not satisfactory, and the effective three-body interaction has played an essential role in understanding the systems.
In this study, we investigate a few-body system comprising of ultracold bosons tightly confined in a deep optical lattice site, which is effectively described as zero-dimensional bosons.
By combining an occupancy-resolving high-resolution laser spectroscopy with an inter-orbital Feshbach resonance controlling the bare two-body interaction over a wide range, 
we experimentally reveal the behaviors of few-atom systems in a strongly-interacting regime.
Our results, for which perturbative calculations do not provide proper explanations, serve as a valuable and precise benchmark for theoretical approaches to strongly-interacting few-body systems.
As one important illustration, we obtain a clear signature of an effective four-body interaction, evidenced by the binding energies of four and more atoms.
This work is an important step for our deeper understanding of strongly-interacting few-body systems.

\end{abstract}


\maketitle


\section{Introduction}
A few-body quantum system is usually defined as the one which consists of a small number of well-defined constituent particles, often trapped or self-trapped under some confining potential \cite{glockle2012quantum}.
Few-quark states, light nuclear and hadronic systems, few-electron atomic and molecular systems, quantum dots, and trapped cold-atom systems are the typical examples, which together with universal aspects across them have recently been studied intensively \cite{NIELSEN2001373,RevModPhys.76.215,BRAATEN2006259,Blume_2012,RevModPhys.85.197,RevModPhys.89.035006,D’Incao_2018}. 
In these few-body systems, a two-body interaction or force between constituent quantum particles, usually described by scattering lengths at a low collision energy, plays a dominant role in the quantitative description of the system.\\
\indent
In nuclear physics, for instance, the details of the two-body nucleon-nucleon interaction via one-boson exchange potential are now rather well established \cite{cdb,AV18,nijm}.
Several important observations, however, cannot be explained in terms of the simple pairwise interaction, and the effective three-body force between nucleons has been introduced to resolve the discrepancies.
Illustrative examples involve the binding energies of light nuclei like triton ($^{3}$H) \cite{PhysRevC.33.1740,PhysRevC.65.054003} and $^{3,6,8}$He \cite{PhysRevC.66.044310}, the deuteron-nucleon scattering cross section \cite{wit98,sakai2000,sekiguchi2002,PhysRevLett.95.162301}, the equation of state of nuclear matters like a massive neutron star \cite{PhysRevC.58.1804} and the anomaly on the limit of neutron-rich nuclei, namely the neutron drip line \cite{PhysRevLett.105.032501}.
Here, the most well-known microscopic mechanism of the effective three-body force is known as the Fujita-Miyazawa type \cite{10.1143/PTP.17.360}, which is induced by the virtual $\Delta$ excitations via the two-body nuclear force due to pion exchange.\\
\indent
Recent progress of ultracold atoms has also provided a useful platform for studying few-body atomic systems.
High controllability and a rich variety of powerful spectroscopic techniques of this cold-atom system enables the investigation of, for example, a novel phenomenon of the Efimov effect in which an effective three-body interaction emerges via resonant two-body interactions, resulting in a series of universal weakly-bound states characterized by a three-body parameter \cite{BRAATEN2007120,annurev:/content/journals/10.1146/annurev.nucl.012809.104439,Ferlaino2011,Naidon_2017}. 
Similarly, four- and more atom bound states are observed and explained by using a three-body parameter or potential \cite{PhysRevLett.102.140401,zenesini2013resonant}. 
Another important example of few-body atomic system is a well-localized few atom system tightly confined in each of an optical lattice potential, and
the effective three-body interaction were extensively studied with various methods such as quantum phase collapse and revival technique \cite{Will2010}, lattice modulation spectroscopy \cite{PhysRevLett.107.175301}, and high-resolution clock spectroscopy both for bosons \cite{Campbell2006,PhysRevLett.110.173201,Franchi_2017} and fermions \cite{Goban2018}.
Such progress on the experimental side has stimulated further development on the theoretical side, including exact solutions to two atoms in a harmonic trap \cite{Busch1998,PhysRevA.71.050701,PhysRevA.74.022712,Liang_2008,PhysRevA.101.053624}, perturbative calculations of effective multi-body interactions \cite{Johnson_2009,Johnson_2012,PhysRevA.83.063609}, and analytical and numerical approaches to strongly-interacting few-body systems \cite{NIELSEN2001373,RevModPhys.76.215,BRAATEN2006259,Blume_2012,RevModPhys.85.197,RevModPhys.89.035006,D’Incao_2018}. 

The emergence of the effective multi-body interactions is understood in the physical picture based on the microscopic description by integrating out the degrees of freedom of virtual excitations to higher vibrational states.
Here, we can find an analogy with the effective three-body force of Fujita-Miyazawa type in nuclear physics, where nucleons, a bare two-body nuclear force, and the $\Delta$ excitations correspond to the atoms in the vibrational ground state of a harmonic trap, the bare two-body interatomic force, and the vibrational excitations of the atoms, respectively.

A few-body quantum system in a strongly-interacting regime beyond the perturbation is in general by no means trivial.  
Note that, already discussed in a pioneering theoretical work \cite{Busch1998}, the strong interaction of the two-body system appears as the nonlinearity of the interaction energy with respect to the scattering length $a_s$ when $a_s$ is comparable to or larger than the confining harmonic oscillator length, and was observed experimentally (see e.g.~\cite{RevModPhys.80.885} and also~\cite{PhysRevLett.110.173201, PhysRevLett.113.120402, scazza2014observation}). 
In contrast, beyond a two-body problem, a few-body quantum system in a strongly-interacting regime has not been assessed, and revealing the behavior will give insights into the nuclear systems where the non-perturbative treatments are generally quite difficult.

In this study, we reveal the novel behavior of the strongly-interacting few-body bosonic atoms, beyond the perturbative treatment of the bare two-body interaction in previous works.
As important illustration of encapsulating the strong interactions of this system, 
we find an interesting feature of the effective three-body interaction where the perturbative calculation up to the second order does not reproduce any of the experimental data even qualitatively, and the third-order result, on the other hand, can capture the tendency of the experimental data.
In addition, we obtain a clear signature of an effective four-body interaction, evidenced by the binding energies of four and more atoms in a lattice site at the strongly-interacting regime.
This is achieved by exploiting the advantageous features of cold-atom optical-lattice quantum simulator, combining an occupancy-resolving high-resolution laser spectroscopy for well-defined few-body atomic systems prepared in a three-dimensional (3D) optical lattice with the precise control of the two-body interaction over a wide range via an inter-orbital Feshbach resonance.
Our results, for which perturbative calculations do not provide proper explanations, serve as a valuable and precise benchmark for theoretical approaches to few-body systems with strong interactions well beyond the regime where the perturbation theory applies.

Effective multi-body interactions have also been of considerable interest in different contexts of the study of localized spin models.
Effective multi-spin interactions, arising from higher-order processes of pairwise exchange interactions, were theoretically proposed to explain the magnetic behaviors in solid helium \cite{RevModPhys.55.1} and some organic compounds \cite{PhysRevB.72.045105}.
In addition, similar effective multi-spin interactions have been recently engineered in artificial systems of a chain of ions in an ion trap \cite{katz2023demonstration} and localized spins loaded into an optical lattice with unit filling \cite{dai2017four}, which would play an important role in quantum computation and quantum information.
Moreover, an effective three-body interaction for photon is successfully implemented by exploiting Rydberg interactions \cite{PhysRevLett.121.103601}. 

Note that, in addition to the effective multi-body interactions, {\it fundamental} multi-body interactions emerge due to the self-interaction of gluons in the microscopic description of quantum chromodynamics (QCD), and multi-pion interactions naturally lead to the fundamental multi-body interactions also for nucleons \cite{RevModPhys.81.1773}. 
Since such self-interaction is absent for photon, multi-body interactions in atomic and molecular systems only emerge in the low-energy effective description by tracing out the high-energy degrees of freedom involved in the microscopic description.

\section{Harmonically trapped bosons}
Here, we present theoretical descriptions of ultracold bosons tightly confined in a harmonic trap.
There are microscopic and effective descriptions, both of which should give the same results.
Note that the hopping between neighboring sites is negligible for a sufficiently deep optical lattice potential, and the trapping potential for atoms in each lattice site is well approximated by a harmonic potential.

{\it Microscopic description.{\bf \textemdash}}
The system consisting of multiple bosons in the ground $g$ and excited $e$ states is described by the second-quantized Hamiltonian, on the one hand,
\begin{align}\label{eq:microscopic}
\hat{H} &= \int\!d\r\Biggl[\sum_{\sigma=g,e}
\hat\phi_\sigma^\+(\r)\left(-\frac{\hbar^2}{2m}\frac{\d^2}{\d\r^2}
+ \frac{m}{2}\omega^2\r^2\right)\hat\phi_\sigma(\r) \notag\\
 &\qquad + \sum_{\sigma,\sigma'}\frac{g_{\sigma\sigma'}}{2}\hat\phi_\sigma^\+(\r)
\hat\phi_{\sigma'}^\+(\r)\hat\phi_{\sigma'}(\r)\hat\phi_\sigma(\r)\Biggr],
\end{align}
where $m$ is the mass of the boson, $g_{\sigma\sigma'}=4\pi \hbar^2 a_{\sigma\sigma'}/m+O(a^2)$ is the coupling constant characterizing the bare two-body interaction via a scattering length $a_{\sigma\sigma'}$, $\omega$ is the trap angular frequency which is now assumed to be isotropic and equal before and after the excitation, and $\hat\phi_\sigma^\+(\r)$ and $\hat\phi_\sigma(\r)$ are the creation and annihilation operators of bosons for the state $\sigma=g,e$, respectively.
While only the two-body interaction term exists in Eq.~(\ref{eq:microscopic}), we can evaluate, in principle, any physical quantities including the binding energies of multiple bosons with this Hamiltonian.

{\it Effective description.{\bf \textemdash}}
On the other hand, we also have another useful low-energy effective description by regarding the system as zero-dimensional bosons where we do not consider their spatial degrees of freedom.
This description is justified for the case where the energy scale determined by the trap frequency is largest among others like the atom temperature and chemical potential.
The effective zero-dimensional Hamiltonian up to five atoms including up to one excited-state atom reads
\begin{align}\label{eq:effective}
\hat{H}_\eff &= U_g\hat{b}_g^\+\hat{b}_g + U_e\hat{b}_e^\+\hat{b}_e
+ \frac{U_{gg}}{2!}\hat{b}_g^{\+2}\hat{b}_g^2
+ U_{eg}\hat{b}_e^\+\hat{b}_e\hat{b}_g^\+\hat{b}_g \notag\\
&\quad + \frac{U_{ggg}}{3!}\hat{b}_g^{\+3}\hat{b}_g^3
+ \frac{U_{egg}}{2!}\hat{b}_e^\+\hat{b}_e\hat{b}_g^{\+2}\hat{b}_g^2
+ \frac{U_{gggg}}{4!}\hat{b}_g^{\+4}\hat{b}_g^4 \notag\\
&\quad + \frac{U_{eggg}}{3!}\hat{b}_e^\+\hat{b}_e\hat{b}_g^{\+3}\hat{b}_g^3
+ \frac{U_{ggggg}}{5!}\hat{b}_g^{\+5}\hat{b}_g^5
+ \frac{U_{egggg}}{4!}\hat{b}_e^\+\hat{b}_e\hat{b}_g^{\+4}\hat{b}_g^4,
\end{align}
where $\hat{b}_\sigma^\+$ and $\hat{b}_\sigma$ are the creation and annihilation operators of bosons in the vibrational ground state of a harmonic trap, respectively.
Effective multi-body interaction energies entering the effective zero-dimensional Hamiltonian in Eq.~(\ref{eq:effective}) can be determined by matching the energies of $N$ atoms with those computed from Eq.~(\ref{eq:microscopic}).

\section{Effective multi-body interactions with perturbation theory} \label{Effective multi-body interactions with perturbation theory}
Before the discussion of the experiment, we here present the perturbation theory of the effective multi-body interactions studied in this work, the results of which are later compared with the experimental observations to illustrate the novel features of the strongly-interacting regime of the multi-body interactions. 
The system consisting of multiple $^{174}$Yb atoms in the ground ${^1}S_0$ and excited ${^3}P_2$ states is described by Eq.~(\ref{eq:microscopic}).
The trap angular frequency $\omega$ is set to be $\omega_g=2\pi\times35.9$~kHz if all $N$ atoms are in the ground state, but it is assumed to be an average value of
\begin{align}
\omega_{eg^{N-1}} &= \frac13\sum_{i=x,y,z}\frac{(N-1)\omega_{gi}+\omega_{ei}}{N} \\
&= 2\pi \times
\begin{cases}
\,39.0\ \mathrm{kHz} \quad (N=2) \\
\,38.0\ \mathrm{kHz} \quad (N=3) \\
\,37.5\ \mathrm{kHz} \quad (N=4)
\end{cases}
\end{align}
under the equal frequency approximation if one of $N$ atoms is excited.
The corresponding harmonic oscillator lengths are denoted by $l_g=\sqrt{\hbar/(m\omega_g)}$ and $l_{eg^{N-1}}=\sqrt{\hbar/(m\omega_{eg^{N-1}})}$, respectively.

The procedure to determine the effective multi-body interaction energies with the perturbation theory is elaborated on in Ref.~\cite{Johnson_2012}, whereas it has to be extended to the case of unequal scattering lengths $a_{gg}\neq a_{eg}$ as in the experimental situation of this study.
The energy of $N$ atoms all in the ground state is provided by
\begin{align}
E_{g^N} = \sum_{n=1}^N\binom{N}{n}U_{g^n}.
\end{align}
By computing the left-hand side up to the third order in scattering lengths, the effective three- and four-body interaction energies are found to be
\begin{align}\label{eq:Uggg}
\frac{U_{ggg}}{\hbar\omega_g}
&= -6\alpha_3^{(2)}\left(\frac{a_{gg}}{l_g}\right)^2
+ 12\beta_3^{(3)}\left(\frac{a_{gg}}{l_g}\right)^3
+ 12\alpha_3^{(3)}\left(\frac{a_{gg}}{l_g}\right)^3 \notag\\
&\quad - 6\alpha_{4,3}^{(3)}\left(\frac{a_{gg}}{l_g}\right)^3 
- 18\alpha_5^{(3)}\left(\frac{a_{gg}}{l_g}\right)^3 + O(a^4)
\end{align}
and
\begin{align}\label{eq:Ugggg}
\frac{U_{gggg}}{\hbar\omega_g}
&= 48\alpha_{4,1}^{(3)}\left(\frac{a_{gg}}{l_g}\right)^3
+ 48\alpha_{4,2}^{(3)}\left(\frac{a_{gg}}{l_g}\right)^3 \notag\\
&\quad - 72\alpha_5^{(3)}\left(\frac{a_{gg}}{l_g}\right)^3 + O(a^4),
\end{align}
respectively.
Here, the coefficients introduced in Ref.~\cite{Johnson_2012} read
\begin{align}
\alpha_3^{(2)} &= 0.142626,\ 
\beta_3^{(3)} = -0.0349197,\ 
\alpha_3^{(3)} = 0.56494(1), \notag\\
\alpha_{4,1}^{(3)} &= 0.077465,\ 
\alpha_{4,2}^{(3)} = 0.051099,\ 
\alpha_{4,3}^{(3)} = 0.438946, \notag\\
\alpha_5^{(3)} &= 0.051916
\end{align}
with the regularization applied to $\beta_3^{(3)}$, which lead to
\begin{align}
U_{ggg} = 2\pi\hbar \times [\underbrace{-0.584}_{O(a^2)}{}
+ {}\underbrace{0.263}_{O(a^3)}{} + O(a^4)]\ \mathrm{kHz}
\end{align}
and
\begin{align}
U_{gggg} = 2\pi\hbar \times [\underbrace{0.229}_{O(a^3)}{} + O(a^4)]\ \mathrm{kHz}
\end{align}
for $a_{gg}=5.55$~nm.

Similarly, the energy of $N$ atoms with one of them being excited is provided by
\begin{align} \label{eq:E_egN-1}
E_{eg^{N-1}} = E_{g^{N-1}} + \sum_{n=0}^{N-1}\binom{N-1}{n}U_{eg^n}.
\end{align}
By computing the left-hand side up to the third order in scattering lengths, the effective three- and four-body interaction energies are found and their explicit formulae are provided by Eqs.~(\ref{eq:Uegg}) and (\ref{eq:Ueggg}) of Appendix~\ref{App:Explicit formulae}.
We note that the effective range does not appear up to the order we are working on \cite{Johnson_2012}.

Instead of $a_{gg}$ and $a_{eg}$ as in Eqs.~(\ref{eq:Uggg}), (\ref{eq:Ugggg}), (\ref{eq:Uegg}), and (\ref{eq:Ueggg}), one may adopt $U_{gg}$ and $U_{eg}$ to parametrize the interactions between two atoms, which provide stabler extrapolations of perturbation because $U_{eg}$ is finite even when $a_{eg}$ diverges at the Feshbach resonance.
The effective three-body and four-body interaction energies in the expansions over $U_{gg}$ and $U_{eg}$ up to their third order are obtained by substituting
\begin{align}
\frac{a_{gg}}{l_g} &= \frac{\Gamma\!\left(-\frac12-\frac{U_{gg}}{2\hbar\omega_g}\right)}
{\sqrt2\,\Gamma\!\left(-\frac{U_{gg}}{2\hbar\omega_g}\right)} \notag\\
&= \sqrt{\frac\pi2}\frac{U_{gg}}{\hbar\omega_g}
- \sqrt{\frac\pi2}(1-\ln2)\left(\frac{U_{gg}}{\hbar\omega_g}\right)^2 + O(U^3) \label{eq:agg_lg}
\end{align}
and
\begin{align}
\frac{a_{eg}}{l_{eg}} &= \frac{\Gamma\!\left(-\frac12-\frac{U_{eg}}{2\hbar\omega_{eg}}\right)}
{\sqrt2\,\Gamma\!\left(-\frac{U_{eg}}{2\hbar\omega_{eg}}\right)} \notag\\
&= \sqrt{\frac\pi2}\frac{U_{eg}}{\hbar\omega_{eg}}
- \sqrt{\frac\pi2}(1-\ln2)\left(\frac{U_{eg}}{\hbar\omega_{eg}}\right)^2 + O(U^3) \label{eq:aeg_leg}
\end{align}
into Eqs.~(\ref{eq:Uggg}), (\ref{eq:Ugggg}), (\ref{eq:Uegg}), and (\ref{eq:Ueggg}).
We find
\begin{align}
\label{eq:Uggg_Uexpansion}
U_{ggg} = 2\pi\hbar \times [\underbrace{-0.612}_{O(U^2)}{}
+ {}\underbrace{0.324}_{O(U^3)}{} + O(U^4)]\ \mathrm{kHz}
\end{align}
and
\begin{align}
\label{eq:Ugggg_Uexpansion}
U_{gggg} = 2\pi\hbar \times [\underbrace{0.245}_{O(U^3)}{} + O(U^4)]\ \mathrm{kHz}
\end{align}
for $U_{gg}=2\pi\hbar\times4.04$ kHz, whereas the explicit formulae of $U_{egg}$ and $U_{eggg}$ are provided by Eqs.~(\ref{eq:Uegg_Uexpansion}) and (\ref{eq:Ueggg_Uexpansion}) of Appendix~\ref{App:Explicit formulae}.
The resulting $\nu_3-\nu_1=[2(U_{eg}-U_{gg})+U_{egg}-U_{ggg}]/(2\pi\hbar)$ as a function of $\nu_2-\nu_1=(U_{eg}-U_{gg})/(2\pi\hbar)$ and $\nu_4-\nu_1=[3(U_{eg}-U_{gg}+U_{egg}-U_{ggg})+U_{eggg}-U_{gggg}]/(2\pi\hbar)$ as a function of $\nu_3-\nu_2=(U_{eg}-U_{gg}+U_{egg}-U_{ggg})/(2\pi\hbar)$ under the variation of $U_{eg}$ in the equal frequency approximation are plotted in Figs.~\ref{Fig.3BF}(a,b) and \ref{Fig.4BF}(a,b), respectively, which are compared with the experimental data later in Sec.~\ref{Experimental Results}.

\section{Experimental setup}
{\it Occupancy-resolving spectroscopy.{\bf \textemdash}}
\begin{figure}[b]
  \includegraphics[width=8.6cm]{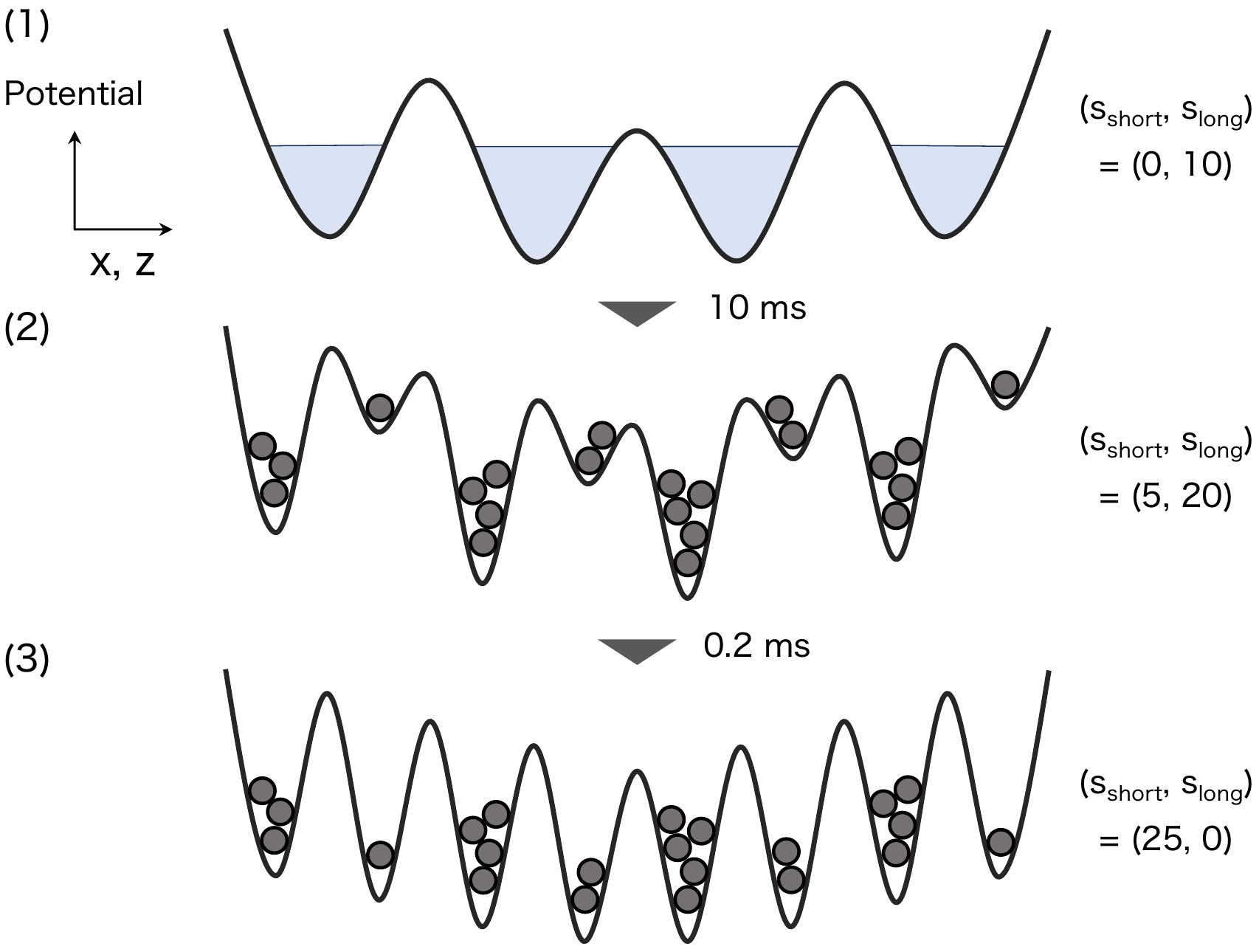}
  \caption{\label{Fig.5} {\bf Schematic illustration of atom loading schemes using optical superlattice.} 
  $s_{\text{short}}$ and $s_{\text{long}}$ are the potential depth of the short lattice (wavelength $\lambda_{\text{short}}=\ $532~nm) and the long lattice ($\lambda_{\text{long}}=\ $1064~nm) normalized by the recoil energy for each wavelength, respectively.
    (1) First we load a $^{174}$Yb BEC to a relatively shallow long lattice along the $x$- and $z$-axes as well as the tight short lattice along the $y$-axis.
    (2) Next, we ramp up a short lattice in 10 ms and trap the atoms in an optical superlattice in a staggered phase, where the atoms are localized at each lattice site.
    (3) Finally, we quickly ramp up the short lattice in 0.2 ms, while we ramp down the long lattice at the same time, where we use the atom-number-projection technique \cite{PhysRevA.99.033609} to increase the number of sites with large atomic occupancy.
    }
\end{figure}
Our experiment starts with the preparation of $^{174}$Yb Bose-Einstein condensate (BEC) by evaporative cooling, and then loads it finally into a 3D cubic optical lattice at the potential depth for each axis of the 3D optical lattice of $25\,E_R$, where $E_R=\hbar^2k_L^2/(2m)$ is the recoil energy of the optical lattice, $k_L=2\pi/\lambda$ with $\lambda=532$~nm is the wave number of the laser for the optical lattice, and $\hbar$ is the Planck constant divided by 2$\pi$.
At this condition, the hopping between neighboring sites is negligible compared with the typical measurement time of 1 ms, and thus this system can be regarded as an ensemble of various isolated quantum few-body systems with a fixed number of particles at each site.

In this study, in order to prepare sites with a large number of occupied atoms up to five, a staggered-phase optical superlattice is used in the initial loading to the optical lattice.
As is shown in Fig.~\ref{Fig.5}, the atoms are first loaded to a shallow long lattice along the $x$- and $z$-axes where the atoms are delocalized over lattice sites.
This is followed by ramping up of a short lattice, resulting in the trapping of the atoms in an optical superlattice in a staggered phase, where the atoms are localized at each lattice site.
Note that the potential difference between adjacent sites in a unit cell of an optical superlattice helps to create sites with large occupancy.
Finally, in order to obtain a large number of sites with large atomic occupancy with a tight confinement, we apply the atom-number-projection technique \cite{PhysRevA.99.033609} where we quickly ramp up the short lattice, while we ramp down the long lattice at the same time (see Appendix~\ref{App:sample} for the detail of the scheme).
\begin{figure}
  \includegraphics[width=8.6cm]{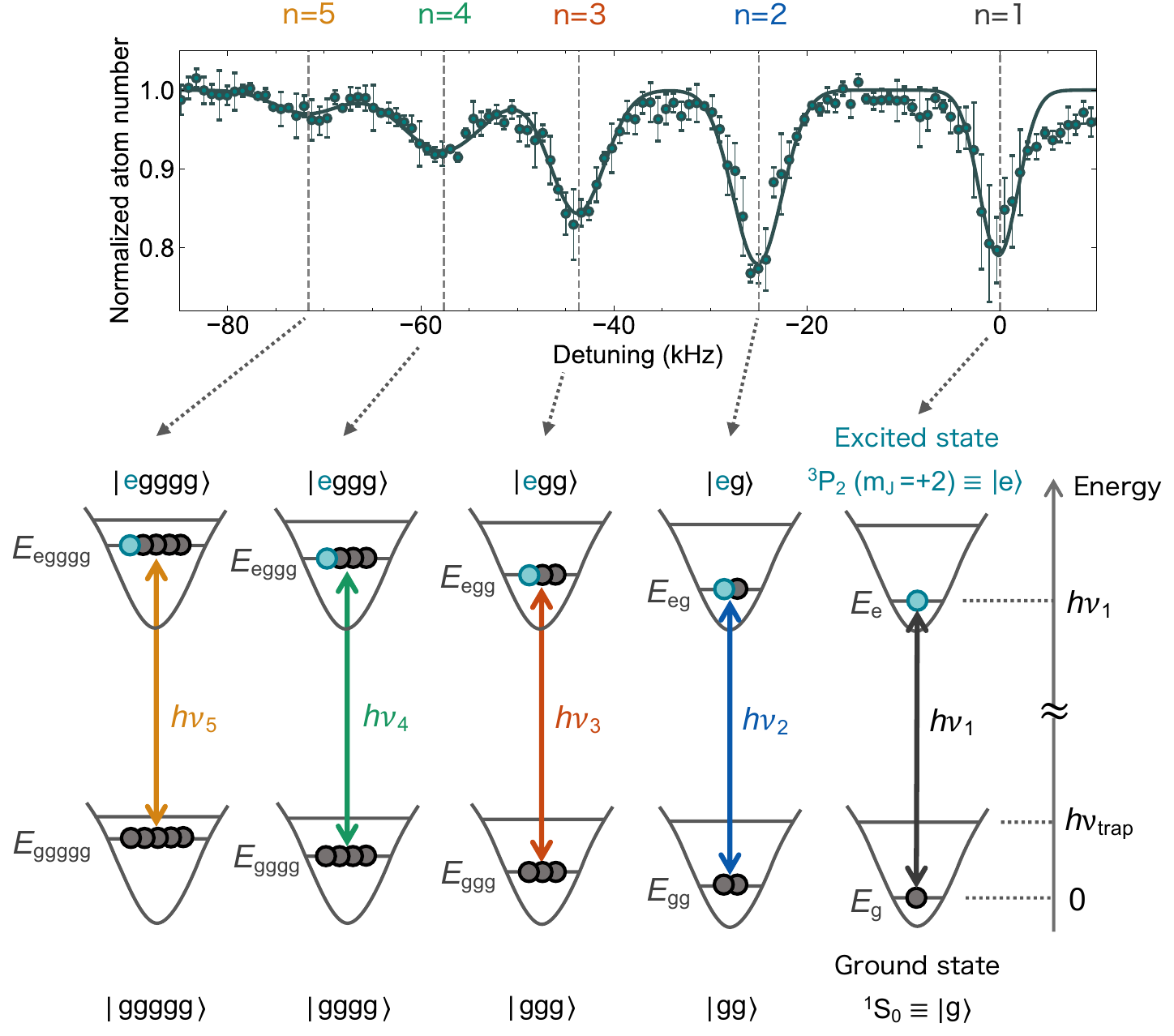}
  \caption{\label{Fig.1} {\bf Occupancy-resolving spectroscopy of $^{174}$Yb atoms in a three-dimensional optical lattice.}
  (Upper) Typical spectroscopic result for the ${^1}S_0\leftrightarrow{^3}P_2\ (m_J=+2)$ transition taken at a magnetic field strength of 232~mG, where the origin of the horizontal frequency axis is the resonance frequency of the singly occupied site.
  The error bars show the standard deviation of the independent four scans.
  The solid line denotes the result of the fit with the multi-Gaussian function.
  Note that the tail of $n=1$ appears on the positive frequency side above about +5~kHz, which is considered to be derived from the excitation of center-of-mass motion for $n=3$ sites (see also Appendix~\ref{App:Analysis}).
  The number of atoms corresponds to the number of ground-state atoms remaining in the crossed dipole trap after the lattice ramp-down sequence, normalized by the number of offset atoms in the fit result.
  The vertical dashed lines represent the resonance positions for the atom-filling, or occupancy of $n=1, 2, ..., 5$.
  (Lower) Schematic diagram of an isolated quantum few-body system realized at each site of an optical lattice.
  Before the laser excitation, the atoms are prepared in a few-body state $\ket{gg\cdots g}$ with the corresponding energy of $E_{gg\cdots g}$.
  By the laser excitation, only one of the atoms in the ground state is excited to the excited state due to the collision blockade effect, resulting in the few-body state of $\ket{eg\cdots g}$ with the energy $E_{eg\cdots g}$.
  The difference in the energies before and after the excitation $E_{eg\cdots g}-E_{gg\cdots g}$ corresponds to the resonance frequency $\nu_n$ for $n$-filling.
  The filling-dependent energy difference leads to a shift of the resonance position from the singly occupied resonance frequency $\nu_1$.
}
\end{figure}
%

After loading BEC into an optical lattice, high-resolution laser spectroscopy of the ultra-narrow magnetic quadrupole ${^1}S_0\leftrightarrow{^3}P_2$ transition is performed.
First, a laser pulse resonant to the ${^1}S_0\leftrightarrow{^3}P_2~(m_J=+2)$ transition, whose wavelength is 507~nm, is irradiated for 1.0~ms.
The optical lattice is then adiabatically lowered to transfer atoms into an optical dipole trap, and absorption imaging is performed to measure the number of atoms remaining in the ground ${^1}S_0$ state.
A typical spectrum obtained in this manner is shown in Fig.~\ref{Fig.1} (upper), where the signals corresponding to different numbers of atom occupancy $n$ at each site of the optical lattice are observed as resonant atom losses.
Here, the resonances corresponding to the singly occupied sites are identified from the spectra with the reduced atom number, and the corresponding resonance frequency is set to zero
in Fig.~\ref{Fig.1} (upper).\\
\indent
From the occupancy-resolved spectrum in Fig.~\ref{Fig.1} (upper), we can access binding energies of the few-body quantum systems.
In fact, as depicted in Fig.~\ref{Fig.1} (lower), the resonance frequency for each occupancy number $n$ corresponds to the difference between the energies before and after the excitation of one atom from the ground state ${^1}S_0$ to the metastable excited state ${^3}P_2~(m_J=+2)$.

\section{Experimental Results} \label{Experimental Results}
{\it Variation of resonance-frequency shifts via a Feshabch resonance.{\bf \textemdash}}
\begin{figure*}
  \includegraphics[width=17.2cm]{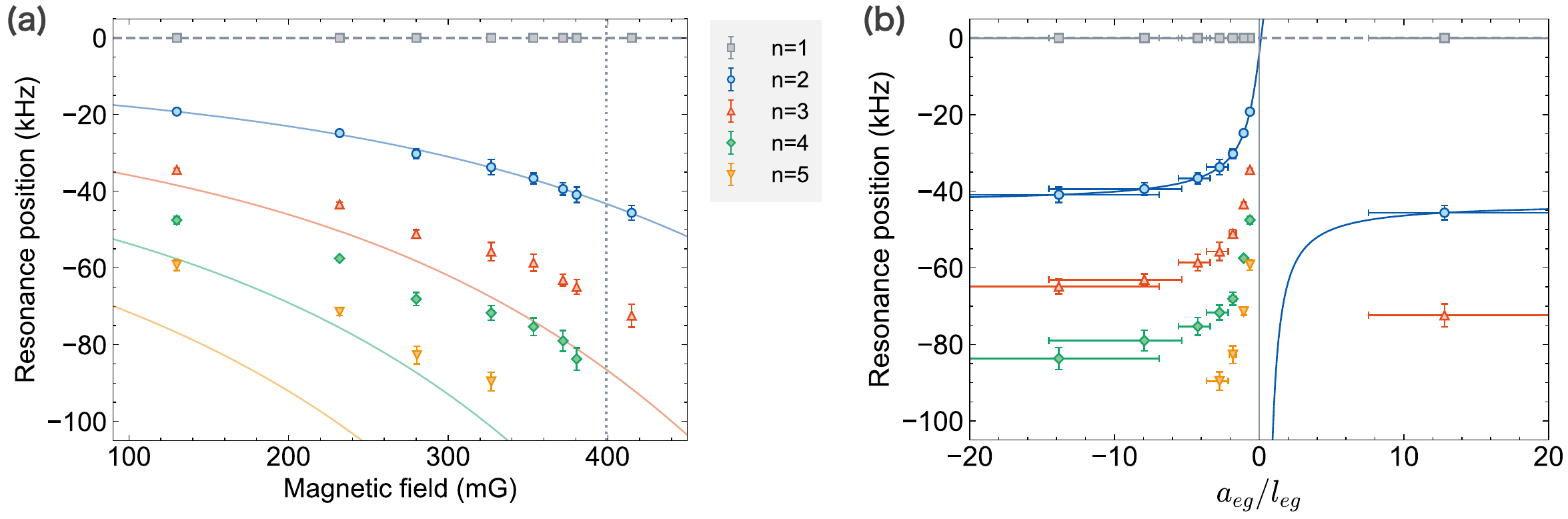}
  \caption{\label{Fig.2} {\bf Variation of resonance positions for $n$-occupancy sites via a Feshbach resonance.}
  {\bf a,} Dependence on a magnetic field.
  The resonance positions corresponding to the $n=1$ (gray square), $n=2$ (blue circle), $n=3$ (red triangle), $n=4$ (green diamond), and $n=5$ (orange inverted triangle) sites obtained from the spectra at each magnetic field strength are plotted.
  Note that the origin of the vertical frequency axis is taken as the resonance frequency of the single occupancy (horizontal dashed line).
  The error bars show the standard deviation of the resonance positions obtained from at least three independent scans, including errors associated with the correction for long-term laser frequency drift (see Appendix~\ref{App:Analysis}).
  The vertical dotted line denotes the Feshbach resonance point between ${^1}S_0\leftrightarrow{^3}P_2\ (m_J=+2)$ atoms at a magnetic field strength of 398 mG, obtained from the measured $n=2$ resonance positions (see Appendix~\ref{App:Scattering}).
  Note that calibration of the magnetic field is done by observing the resonance position of ${^1}S_0\leftrightarrow{^3}P_2\ (m_J=\pm2)$, and the accuracy of the magnetic field is about few mG.
  The solid curves show calculated resonance positions including only the two-body interaction terms, where the solid curve at $n=2$ (blue) shows a result of the theoretical calculation.
  The solid curves at $n=3, 4$ and $5$ (red, green and orange) show a result of the solid curve at $n=2$ (blue) multiplied by 2, 3 and 4, respectively, which for $n$ occupancies are explicitly given by $(n-1)(U_{eg}-U_{gg})$ (see Appendices~\ref{App:Scattering} and \ref{App:Derivation}).
  We note that the relatively large deviation of the measured resonance positions from the solid curves, even in a regime where $a_{eg}$ is relatively small, is due to neglecting the contributions of effective three- and four-body forces.
  Note that the experimental data in the perturbative regime in Fig.~\ref{Fig.3BF}(a) and Fig.~\ref{Fig.4BF}(a) are close to the calculation, where the effective three- and four-body forces are perturbatively taken into consideration.
  {\bf b,} Dependence on the scattering length $a_{eg}$.
  The legends are common to those of {\bf a}.
  The horizontal axis represents the scattering length $a_{eg}$ for atoms in the ground ${^1}S_0$ and excited ${^3}P_2\ (m_J=+2)$ states, divided by the harmonic oscillator length $l_{eg}=39$~nm corresponding to the optical lattice depth of $25\,E_R$.
  The blue solid curve at the $n=2$ plots shows a result of the theoretical calculation.
  The horizontal error bars are evaluated for each field strength from the error in the determination of the scattering length (see Appendix~\ref{App:Scattering}).
  Note that the energy shift of about $-3$~kHz for $a_{eg}=0$ in {\bf b} is due to the non-zero value of the interaction energy between the ground-state atoms ($U_{gg}$).
}
\end{figure*}
In order to enter the strongly-interacting regime beyond the perturbative treatment of the interatomic interaction, we combine the inter-orbital Feshbach resonance with the occupancy-resolving high-resolution spectroscopy.
Here, the scattering length $a_{eg}$ between the $^{174}$Yb atoms in the ground ${^1}S_0$ and excited ${^3}P_2~(m_J=+2)$ states is controlled with a magnetic field via an anisotropy-induced Feshbach resonance \cite{PhysRevLett.110.173201}.
Figure~\ref{Fig.2} shows the measured frequency shift of the resonance position of the multiply occupied site with respect to the resonance frequency of the singly occupied site, as a function of (a) a magnetic field and (b) the scattering length which is determined from the $n=2$ data (see Appendix~\ref{App:Scattering} for the details of the calculation).
As the magnetic field approaches the value at the Feshbach resonance point indicated by the dotted line (see Appendix~\ref{App:Scattering}), the frequency shift also increases, reflecting the larger absolute values of the scattering length $a_{eg}$.
We note that the resonance signals corresponding to the multiple occupancies, $n=5$ in particular, become weaker around the Feshbach resonance, possibly due to the spectral broadening caused by the fast three-body inelastic loss of the atoms in the excited state.
We also note that the three-dimensional lattice is considered as an ideally isolated matrix of zero-dimensional systems even in the strongly interacting regime, since the band widths for the ground as well as higher bands are about one-order-of-magnitude smaller than the band gaps and the interatomic interaction energy.
\\
\indent
The measured spectroscopic shifts for site-occupations at a strongly-interacting regime clearly deviate from those expected withina two-body description (solid curves in Fig.~\ref{Fig.2}(a)).
Note that strong non-linearity of interactions already manifests itself at the two-body level. In fact, the scattering lengths are determined by fitting the observed two-body resonances with the analytical formula given in Ref.~\cite{Busch1998} (see Appendix~\ref{App:Scattering}).
\\
\indent
Since the strong interactions are induced via the inter-orbital Feshbach resonance between $(n-1)$ $\ket{g}$-state bosons and a single $\ket{e}$-state atom, rather than among $n$ $\ket{g}$-state bosons, the injected single $\ket{e}$-state atom introduces the modification of the atomic wavefunctions of the relative motion between the $\ket{e}$-state and other $\ket{g}$-state atoms, resulting in the non-linearity of interactions.
In addition, it is instructive to consider this injected few-body state in the context of the Bose polaron problem \cite{scazza2022repulsive,Grusdt_2025} by regarding a single $\ket{e}$-state atom as an impurity.\\
\indent
{\it Effective multi-body interactions in strongly-interacting regime.{\bf \textemdash}}
As an important example of encapsulating the strong interactions of a confined few-body system, we illustrate the novel behaviors of effective multi-body interactions. 
While the emergence of effective multi-body interactions has been already corroborated in both theoretical and experimental works (e.g.~Refs.~\cite{Johnson_2009}, \cite{Will2010}, and \cite{Goban2018}) as is mentioned in the introductory section, the manner in which strong interactions involve not only modified two-body effective terms but also effective multi-body interactions is by no means trivial.
In fact, the study of the effective multi-body interaction in the strongly-interacting regime was made possible by our experimental access to stronger interactions than those demonstrated so far by exploiting the inter-orbital Feshbach resonance.

\begin{figure}
  \includegraphics[width=8.6cm]{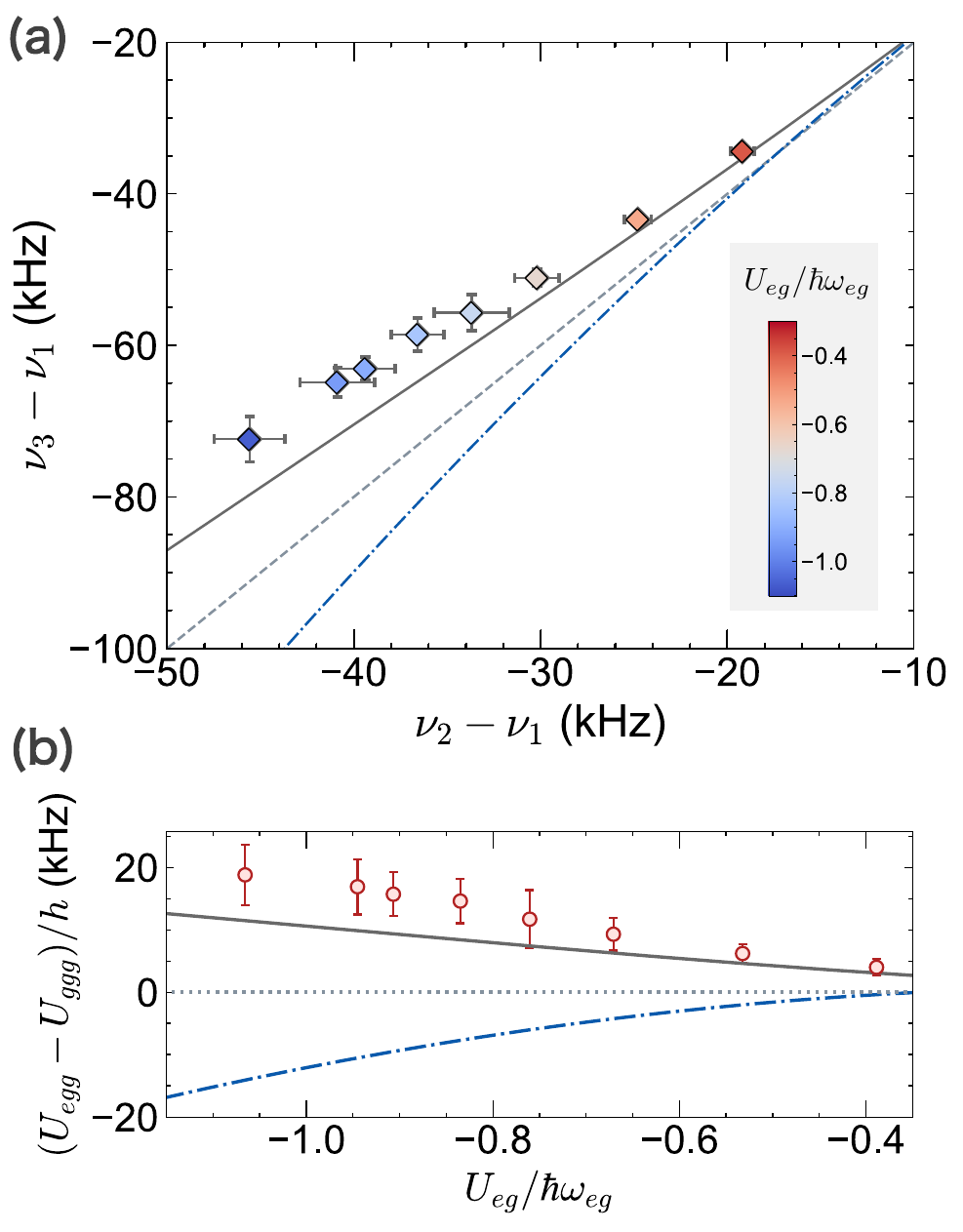}
  \caption{\label{Fig.3BF} {\bf Effective three-body interaction in strongly-interacting regime.}
  {\bf a,} Evidence of an effective three-body interaction in strongly-interacting regime obtained for the $n=3$ resonance.
  The horizontal and vertical axes show the frequency shifts of $n=2$ and $n=3$ with respect to the resonance position of $n=1$, respectively.
  The colorbar shows the two-body interaction energy $U_{eg}$ divided by $\hbar\omega_{eg}$.
  The error bars for each axis are the same as those in Fig.~\ref{Fig.2}.
  The dashed line represents a linear relation with slope 2, which holds between $\nu_3-\nu_1$ and $\nu_2-\nu_1$ when only the effective two-body interactions are incorporated, and the deviation from this line in the plots directly indicates the presence of effective three-body interactions.
  {\bf b,} Amplitude of effective three-body interactions. Plotted here are values of $(U_{egg}-U_{ggg})/h=(\nu_3-\nu_1)-2(\nu_2-\nu_1)$ (Eq.~(\ref{eq:nu3-nu1})) from measured resonance positions of $n=2$ and $3$. 
  The error bars are calculated with the error propagation formula by using the error bars in Fig.~\ref{Fig.3BF}(a).
  In both {\bf a} and {\bf b}, the dashdot and solid curves show the results of the perturbative calculations up to the second and third orders of the two-body interaction energy, respectively.
}
\end{figure}

{\it Effective three-body interaction in strongly-interacting regime.{\bf \textemdash}}
We show the results for an effective three-body interaction in the strongly-interacting regime, which is beyond the perturbative treatment in previous works.
Figure~\ref{Fig.3BF}(a) shows the frequency shifts of $n=2$ and $n=3$ with respect to the resonance position of $n=1$ in the $(\nu_2-\nu_1, \nu_3-\nu_1)$ plane.
Here, in the presence of effective two-body interactions alone, the energy of the three-body system is obtained simply by pairwise interactions, and thus, a linear relationship of 
\begin{align}\label{eq:linear3} 
\nu_3-\nu_1=2(\nu_2-\nu_1)
\end{align}
 holds, which is shown as the dashed line in Fig.~\ref{Fig.3BF}(a) (see Appendix~\ref{App:Derivation}).
Clearly, we find the significant deviation between the experimental data and the dashed line, especially for the large scattering length region, corresponding to the lower left of the figure.

In order to quantitatively discuss the significance of the effective three-body interaction, we perform a chi-square test on the validity of Eq.~(\ref{eq:linear3}), namely, the relation of $y=2x$, where $x$ and $y$ represent the resonance frequency shifts of $n=2$ and $n=3$ from $n=1$, respectively, by evaluating the $\chi^2$ function defined as
\begin{align}
  \chi^2_{\text{3-body}} \coloneqq \sum_i \left[ \left(\frac{x_i-\bar{x}_i}{\sigma_{\bar{x}_i}}\right)^2 + \left(\frac{2x_i-\bar{y}_i}{\sigma_{\bar{y}_i}}\right)^2 \right]. \label{chi_sq_3body}
\end{align}
Here, $i$ is the index representing the data for each magnetic field, $x_i$ are fitting parameters, and $(\bar{x}_i, \sigma_{\bar{x}_i})$ and $(\bar{y}_i, \sigma_{\bar{y}_i})$ are the measured resonance position shifts and their uncertainties for $n=2$ and $n=3$ from $n=1$, respectively.
In this $\chi^2$ function, we search for the fitting parameters that minimize the sum of squares, and obtain the minimum value, $\chi^2_{\text{3-body},\text{min}}=110$ (degree of freedom, $\text{dof}=8$).
This result corresponds to $p\text{-value}=3.33\times10^{-20}$, which corresponds to a significance of $9.2\,\sigma$.
In this way, we find the presence of the three-body interaction with such a large significance,
owing to the Feshbach resonance control of the scattering length.

An interesting feature is found in the comparison of the experimental data with the perturbative calculations described in Sec.~\ref{Effective multi-body interactions with perturbation theory}.
The dashdot and solid curves in Fig.~\ref{Fig.3BF}(a) show the results of perturbative calculations up to the second and third orders of the two-body interaction energy, respectively (see Eqs.~(\ref{eq:Uggg_Uexpansion}) and (\ref{eq:Uegg_Uexpansion})).
Clearly, the second-order perturbation calculation does not reproduce any of the experimental data observed in our experiment on the one hand.
The third-order result, on the other hand, can properly reproduce the data corresponding to the small scattering lengths where the perturbative calculation is justified.
Note that the experimental data are all located on the upper side of the line of Eq.~(\ref{eq:linear3}), which indicates the inequality $U_{egg} > U_{ggg}$ for the effective three-body interactions and the amplitude of the effective three-body interactions $U_{egg}-U_{ggg}$ is plotted in Fig.~\ref{Fig.3BF}(b) as a function of $U_{eg}/\hbar\omega_{eg}$.
This feature is properly captured by the perturbation calculation up to the third order, although the magnitudes of the deviations are not.

%
\begin{figure*}
  \includegraphics[width=17.2cm]{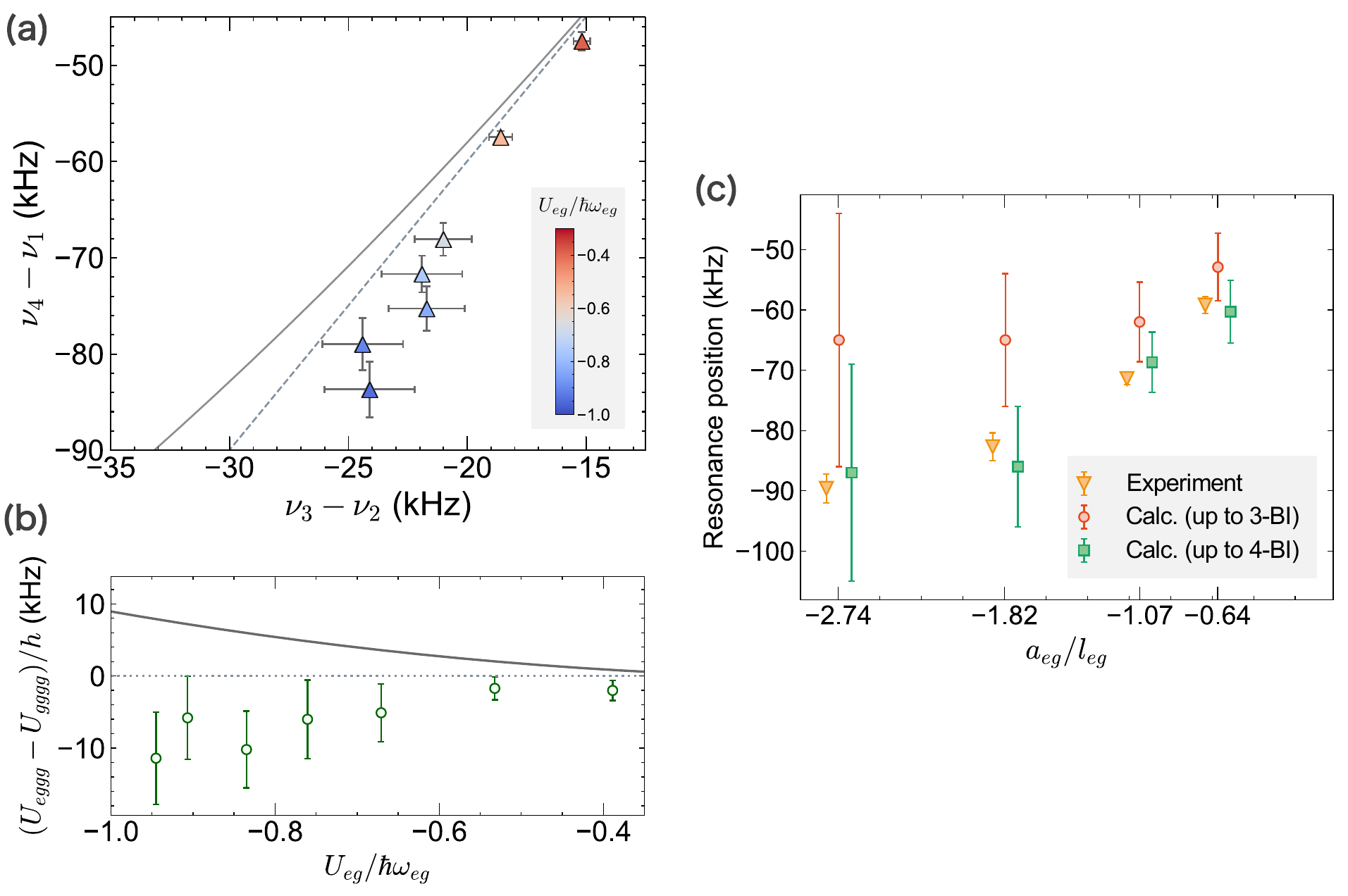}
  \caption{\label{Fig.4BF} {\bf Evidence of an effective four-body interaction.}
  {\bf a,} Evidence of an effective four-body interaction obtained for the $n=4$ resonance.
  The horizontal and vertical axes represent the difference between the $n=2$ and $n=3$ resonance positions and that between $n=1$ and $n=4$, respectively.
  The colorbar shows the two-body interaction energy $U_{eg}$ divided by $\hbar\omega_{eg}$.
  Error bars for the vertical axis are the same as those in Fig.~\ref{Fig.2}, and error bars for the horizontal axis represent the standard deviation obtained in the analysis with respect to the resonance position of $n=2$ (see Appendix~\ref{App:Analysis}).
  The dashed line represents a linear relation with slope 3, which holds between $\nu_4-\nu_1$ and $\nu_3-\nu_2$ when only up to effective three-body interactions are incorporated, and the deviation from this line in the plots is the direct evidence of the presence of effective four-body interactions.
  {\bf b,} Amplitude of effective four-body interactions. Plotted here are values of $(U_{eggg}-U_{gggg})/h=(\nu_4-\nu_1)-3(\nu_3-\nu_2)$ (Eq.~(\ref{eq:nu4-nu1})) from measured resonance positions of $n=2$, $3$, and $4$. 
  The error bars are calculated with the error propagation formula by using the error bars in Fig.~\ref{Fig.4BF}(a).
  In both {\bf a} and {\bf b}, the solid curve shows the result of the perturbative calculation up to the third order of the two-body interaction energy.
  {\bf c,} Contribution of effective four-body interactions to the $n=5$ resonance.
  Plotted here are the measured $n=5$ resonance positions (orange inverted triangles), the calculated $n=5$ resonance positions by incorporating only up to effective three-body interactions (3-BI, red circles), and the calculated $n=5$ resonance positions by incorporating up to effective four-body interactions (4-BI, green squares).
  Note that the plots at each scattering length are slightly shifted horizontally to avoid the overlap of the symbols.
  The error bars for the measured $n=5$ resonance positions are the same as the error bars in Fig.~\ref{Fig.2}, and the error bars in the calculated values are obtained with the error propagation formula by using the error bars for $n=2, 3$, and $4$ resonance positions.
}
\end{figure*}
%
%

{\it Evidence of effective four-body interaction.{\bf \textemdash}}
We provide an evidence regarding the presence of effective four-body interactions.
Figure~\ref{Fig.4BF}(a) shows the frequency difference between the resonance positions of $n=2$ and $n=3$, and that between $n=1$ and $n=4$ at each magnetic field, or scattering length in the $(\nu_3-\nu_2, \nu_4-\nu_1)$ plane.
Here, in the absence of effective four-body interaction, it can be shown that a linear relationship of 
\begin{align}\label{eq:linear4}
\nu_4-\nu_1=3(\nu_3-\nu_2)
\end{align}
holds (see Appendix~\ref{App:Derivation}), which is indicated by the dashed line in Fig.~\ref{Fig.4BF}(a).
Therefore, the breakdown of this linearity is the smoking gun of the effective four-body interaction.
\\
\indent
In Fig.~\ref{Fig.4BF}(a), especially in the large scattering length region
corresponding to the lower left of the figure, a significant deviation from the linear relationship is clearly seen, which is the direct evidence of effective four-body interaction.
In order to quantitatively discuss the significance of the effective four-body interaction, we quantitatively evaluate the deviation from the linear relationship Eq.~(\ref{eq:linear4}). 
Similar to the previous subsection, we perform a chi-square test on the validity of Eq.~(\ref{eq:linear4}), namely, the relation of $z=3v$, where $z$ and $v$ represent the resonance frequency shift of $n=4$ from $n=1$ and that of $n=3$ from $n=2$, respectively, by evaluating the $\chi^2$ function defined as
\begin{align}
  \chi^2_{\text{4-body}} \coloneqq \sum_i \left[ \left(\frac{v_i-\bar{v}_i}{\sigma_{\bar{v}_i}}\right)^2 + \left(\frac{3v_i-\bar{z}_i}{\sigma_{\bar{z}_i}}\right)^2 \right]. \label{chi_sq_4body}
\end{align}
Here, $v_i$ are fitting parameters, $(\bar{v}_i, \sigma_{\bar{v}_i})$ and $(\bar{z}_i, \sigma_{\bar{z}_i})$ are the measured resonance positions and their uncertainties for $n=3$ and $n=4$ with respect to the resonance positions of $n=2$ and $n=1$, respectively, and the other symbols are the same as in Eq.~(\ref{chi_sq_3body}).
We obtain $\chi^2_{\text{4-body},\text{min}}=13.9$ ($\text{dof}=7$) and $p\text{-value}=0.0534$, which corresponds to a significance of $1.9\,\sigma$.
This result indicates that we successfully find the clear experimental evidence of the effective four-body interaction (see also Appendix~\ref{App:Revisiting}).\\
\indent
Note that the plots are all located on the lower side of the line, which shows that the inequality $U_{eggg}<U_{gggg}$ holds for the effective four-body interactions, where $U_{gggg}$ and $U_{eggg}$ represent the interaction energy for the four atoms in the ground state $\ket{g}$ and that for three in the ground state and one in the excited state $\ket{e}$, respectively, and the amplitude of the effective four-body interactions $U_{eggg}-U_{gggg}$ is plotted in Fig.~\ref{Fig.4BF}(b) as a function of $U_{eg}/\hbar\omega_{eg}$.
Since the Feshbach resonance is only induced between the atoms in the ground and excited states and not between the ground state atoms, $U_{eggg}$ should be responsible for the variation of the effective four-body interaction.\\
\indent
Figure~\ref{Fig.4BF}(a,b) also show as a solid curve the result of the perturbative calculation up to the third order of the two-body interaction energy, valid only in a region of relatively small scattering length (see Eqs.~(\ref{eq:Ugggg_Uexpansion}) and (\ref{eq:Ueggg_Uexpansion})).
Clearly, the perturbative calculations do not reproduce the experimental data including the direction of the deviation from the linear relationship Eq.~(\ref{eq:linear4}), indicating the non-perturbative nature of the observed effective four-body interaction.
Note that the possible effects coming from non-harmonic potential terms, as well as the uncertainties due to the magnetic field calibration, $a_{gg}$, the lattice depth calibration, and the trap frequencies, are negligible in our conditions.
These error budgets are relevant only in comparison of the experimental values and the perturbative calculations, and is irrelevant to the discussion on the existence of effective three- and four-body interactions.

\indent
Furthermore, the experimental data of $n=5$ resonance positions also support the presence of the effective four-body interaction.
Figure~\ref{Fig.4BF}(c) shows the measured $n=5$ resonance positions with respect to that of $n=1$ ($(\nu_5-\nu_1)$, inverted triangles), the calculated $n=5$ resonance positions including only effective two- and three-body interactions ($(\nu_5-\nu_1)'$, circles), and the calculated $n=5$ resonance positions including effective two-, three-, and four-body interactions ($(\nu_5-\nu_1)''$, squares).
Here, 
\begin{align}
  (\nu_5-\nu_1)' &= 6(\nu_3-\nu_1)-8(\nu_2-\nu_1), \label{eq:linear5'}\\
  (\nu_5-\nu_1)'' &= 4(\nu_4-\nu_1)-6(\nu_3-\nu_1)+4(\nu_2-\nu_1). \label{eq:linear5''}
\end{align}
See Appendix~\ref{App:Derivation} for the derivation.
To elucidate the contribution of the effective four-body interaction in the binding energy of the $n=5$ sites, we perform a chi-square test. 
The $\chi^2$ functions are defined as
\begin{equation}
  \begin{split}
    (\chi^2_{\text{4-body}})' \coloneqq \sum_i &\Bigg[ \left(\frac{x_i-\bar{x}_i}{\sigma_{\bar{x}_i}}\right)^2+\left(\frac{y_i-\bar{y}_i}{\sigma_{\bar{y}_i}}\right)^2 \\
&+\left(\frac{6y_i-8x_i-\bar{w}_i}{\sigma_{\bar{w}_i}}\right)^2 \Bigg] \label{chi_sq_4body_2}
  \end{split}
\end{equation}
for testing the validity of Eq.~(\ref{eq:linear5'}) which holds in the absence of effective four-body interactions, and as
\begin{equation}
  \begin{split}
    (\chi^2_{\text{4-body}})'' &\coloneqq \sum_i \Bigg[ \left(\frac{x_i-\bar{x}_i}{\sigma_{\bar{x}_i}}\right)^2 + \left(\frac{y_i-\bar{y}_i}{\sigma_{\bar{y}_i}}\right)^2\\ 
    &+ \left(\frac{z_i-\bar{z}_i}{\sigma_{\bar{z}_i}}\right)^2
    +\left(\frac{4z_i-6y_i+4x_i-\bar{w}_i}{\sigma_{\bar{w}_i}}\right)^2 \Bigg] \label{chi_sq_4body_3}
  \end{split}
\end{equation}
for testing the validity of Eq.~(\ref{eq:linear5''}) which holds in the presence of the effective four-body interactions.
Here, $x_i$, $y_i$, and $z_i$ are fitting parameters, $(\bar{w}_i, \sigma_{\bar{w}_i})$ are the measured resonance position shifts and their uncertainties for $n=5$ from $n=1$, and the other symbols are the same as in Eqs.~(\ref{chi_sq_3body}) and (\ref{chi_sq_4body}).
Note that we do not consider effective five-body interactions in both cases, and $6y_i-8x_i$ in the third term of Eq.~(\ref{chi_sq_4body_2}) and $4z_i-6y_i+4x_i$ in the fourth term of Eq.~(\ref{chi_sq_4body_3}) reflect Eqs.~(\ref{eq:linear5'}) and (\ref{eq:linear5''}), respectively.
In these $\chi^2$ functions, we obtain $(\chi^2_{\text{4-body}})'_{\text{min}}=6.71$ ($\text{dof}=4$) and $(\chi^2_{\text{4-body}})''_{\text{min}}=0.46$ ($\text{dof}=4$), which correspond to the $p\text{-values}$ of 0.152 and 0.977, respectively.
This analysis indicates that the measured $n=5$ resonance positions are explained more plausibly by the presence of the effective four-body interaction.
Note that while the absolute magnitude of the effective four-body interaction should increase as the number of occupancies increases, the evidence of the effective four-body interaction for $n=5$ sites is not as clear as that for $n=4$ due to the specific nature of error propagation.
%
\section{CONCLUSIONS AND PROSPECTS}
In this work, we reveal the behaviors of strongly-interacting few-body quantum systems comprising of harmonically confined spinless bosons in a wide range of bare two-body interaction with a high-resolution spectroscopy.
Strong interactions are induced between the $(n-1)$ $\ket{g}$-state bosons and a single $\ket{e}$-state atom, created as a result of the high-resolution laser excitation. 
It is straightforward to apply the same methodology to study strongly-interacting regime of spinful atoms of both fermions and bosons, and also with long-range dipolar interactions provided by magnetic atoms and molecules or Rydberg atoms.
Systematic investigations of these various systems would give important insights into, for example, how the few-body limit can illuminate the quantum many-body problem, by taking larger values of the number of atoms,
which may provide a possible relation to a strongly-interacting bosonic liquid.

\acknowledgments
We thank T.~Takano for his helpful advice regarding the analysis of experimental data, and useful discussions with T.~Nakamura, S.~Endo, T.~Doi, T.~Hyodo, Y.~Kanada, E.~Hiyama, E.~Epelbaum, and K.~Sekiguchi.
This work was supported by the Grant-in-Aid for Scientific Research of JSPS (No.\ JP17H06138, No.\ JP18H05405, No.\ JP18H05228, No.\ JP21H01014, No.\ JP21K03384, No.\ JP22K20356), JST CREST (Nos.\ JPMJCR1673 and JPMJCR23I3), and MEXT Quantum Leap Flagship Program
(MEXT Q-LEAP) Grant No.\ JPMXS0118069021, and JST Moonshot R\&D Grant \ JPMJMS2268 and No.\ JPMJMS2269.
K.H.\ acknowledges the support from the establishment of university fellowships towards the creation of science technology innovation, Grant Number JPMJFS2123.
\appendix
\section{Explicit formulae of the effective multi-body interactions with perturbation theory} \label{App:Explicit formulae}
The energy of $N$ atoms with one of them being excited is provided by Eq.~(\ref{eq:E_egN-1}) of the main text.
By computing its left-hand side up to the third order in scattering lengths, the effective three-body and four-body interaction energies are found to be
\begin{widetext}\begin{align}\label{eq:Uegg}
\frac{U_{egg}}{\hbar\omega_{egg}}
&= -4\alpha_3^{(2)}\left(\frac{a_{gg}}{l_{egg}}\right)
\left(\frac{a_{eg}}{l_{egg}}\right)
- 2\alpha_3^{(2)}\left(\frac{a_{eg}}{l_{egg}}\right)^2 \notag\\
&\quad + 4\beta_3^{(3)}\left(\frac{a_{gg}}{l_{egg}}\right)^2
\left(\frac{a_{eg}}{l_{egg}}\right)
+ 4\beta_3^{(3)}\left(\frac{a_{gg}}{l_{egg}}\right)
\left(\frac{a_{eg}}{l_{egg}}\right)^2
+ 4\beta_3^{(3)}\left(\frac{a_{eg}}{l_{egg}}\right)^3 \notag\\
&\quad + 2\alpha_3^{(3)}\left(\frac{a_{gg}}{l_{egg}}\right)^2
\left(\frac{a_{eg}}{l_{egg}}\right)
+ 8\alpha_3^{(3)}\left(\frac{a_{gg}}{l_{egg}}\right)
\left(\frac{a_{eg}}{l_{egg}}\right)^2
+ 2\alpha_3^{(3)}\left(\frac{a_{eg}}{l_{egg}}\right)^3 \notag\\
&\quad - 2\alpha_{4,3}^{(3)}\left(\frac{a_{gg}}{l_{egg}}\right)^2
\left(\frac{a_{eg}}{l_{egg}}\right)
- 2\alpha_{4,3}^{(3)}\left(\frac{a_{gg}}{l_{egg}}\right)
\left(\frac{a_{eg}}{l_{egg}}\right)^2
- 2\alpha_{4,3}^{(3)}\left(\frac{a_{eg}}{l_{egg}}\right)^3 \notag\\
&\quad - 4\alpha_5^{(3)}\left(\frac{a_{gg}}{l_{egg}}\right)^2
\left(\frac{a_{eg}}{l_{egg}}\right)
- 10\alpha_5^{(3)}\left(\frac{a_{gg}}{l_{egg}}\right)
\left(\frac{a_{eg}}{l_{egg}}\right)^2
- 4\alpha_5^{(3)}\left(\frac{a_{eg}}{l_{egg}}\right)^3 + O(a^4)
\end{align}
and
\begin{align}\label{eq:Ueggg}
\frac{U_{eggg}}{\hbar\omega_{eggg}}
&= 24\alpha_{4,1}^{(3)}\left(\frac{a_{gg}}{l_{eggg}}\right)^2
\left(\frac{a_{eg}}{l_{eggg}}\right)
+ 24\alpha_{4,1}^{(3)}\left(\frac{a_{gg}}{l_{eggg}}\right)
\left(\frac{a_{eg}}{l_{eggg}}\right)^2 \notag\\
&\quad + 30\alpha_{4,2}^{(3)}\left(\frac{a_{gg}}{l_{eggg}}\right)^2
\left(\frac{a_{eg}}{l_{eggg}}\right)
+ 12\alpha_{4,2}^{(3)}\left(\frac{a_{gg}}{l_{eggg}}\right)
\left(\frac{a_{eg}}{l_{eggg}}\right)^2
+ 6\alpha_{4,2}^{(3)}\left(\frac{a_{eg}}{l_{eggg}}\right)^3 \notag\\
&\quad - 42\alpha_5^{(3)}\left(\frac{a_{gg}}{l_{eggg}}\right)^2
\left(\frac{a_{eg}}{l_{eggg}}\right)
- 24\alpha_5^{(3)}\left(\frac{a_{gg}}{l_{eggg}}\right)
\left(\frac{a_{eg}}{l_{eggg}}\right)^2
- 6\alpha_5^{(3)}\left(\frac{a_{eg}}{l_{eggg}}\right)^3 + O(a^4),
\end{align}
respectively.
The effective three-body and four-body interaction energies in the expansions over $U_{gg}$ and $U_{eg}$ up to their third order are obtained by substituting Eqs.~(\ref{eq:agg_lg}) and (\ref{eq:aeg_leg}) of the main text into Eqs.~(\ref{eq:Uegg}) and (\ref{eq:Ueggg}).
We find
\begin{align}
\frac{U_{egg}}{\hbar\omega_{egg}}
&= -\left[4\alpha_3^{(2)}\left(\sqrt{\frac\pi2}\frac{U_{gg}}{\hbar\omega_g}\right)\left(\sqrt{\frac\pi2}\frac{U_{eg}}{\hbar\omega_{eg}}\left(\frac{\omega_g}{\omega_{eg}}\right)^{1/2}\right)
+ 2\alpha_3^{(2)}\left(\sqrt{\frac\pi2}\frac{U_{eg}}{\hbar\omega_{eg}}\left(\frac{\omega_g}{\omega_{eg}}\right)^{1/2}\right)^2\right]\left(\frac{\omega_{egg}}{\omega_g}\right) \notag\\
&\quad - \left[4\beta_3^{(3)}\left(\sqrt{\frac\pi2}\frac{U_{gg}}{\hbar\omega_g}\right)^2\left(\sqrt{\frac\pi2}\frac{U_{eg}}{\hbar\omega_{eg}}\right)\left(\frac{\omega_g}{\omega_{eg}}\right)^{1/2}
+ 4\beta_3^{(3)}\left(\sqrt{\frac\pi2}\frac{U_{gg}}{\hbar\omega_g}\right)\left(\sqrt{\frac\pi2}\frac{U_{eg}}{\hbar\omega_{eg}}\right)^2\left(\frac{\omega_g}{\omega_{eg}}\right)^{1/2}\right. \notag\\
&\qquad\left. + 4\beta_3^{(3)}\left(\sqrt{\frac\pi2}\frac{U_{eg}}{\hbar\omega_{eg}}\right)^3\left(\frac{\omega_g}{\omega_{eg}}\right)\right]\left(\frac{\omega_{egg}}{\omega_g}\right) \notag\\
&\quad + \left[4\beta_3^{(3)}\left(\sqrt{\frac\pi2}\frac{U_{gg}}{\hbar\omega_g}\right)^2\left(\sqrt{\frac\pi2}\frac{U_{eg}}{\hbar\omega_{eg}}\left(\frac{\omega_g}{\omega_{eg}}\right)^{1/2}\right)
+ 4\beta_3^{(3)}\left(\sqrt{\frac\pi2}\frac{U_{gg}}{\hbar\omega_g}\right)\left(\sqrt{\frac\pi2}\frac{U_{eg}}{\hbar\omega_{eg}}\left(\frac{\omega_g}{\omega_{eg}}\right)^{1/2}\right)^2\right. \notag\\
&\qquad + 4\beta_3^{(3)}\left(\sqrt{\frac\pi2}\frac{U_{eg}}{\hbar\omega_{eg}}\left(\frac{\omega_g}{\omega_{eg}}\right)^{1/2}\right)^3
+ 2\alpha_3^{(3)}\left(\sqrt{\frac\pi2}\frac{U_{gg}}{\hbar\omega_g}\right)^2\left(\sqrt{\frac\pi2}\frac{U_{eg}}{\hbar\omega_{eg}}\left(\frac{\omega_g}{\omega_{eg}}\right)^{1/2}\right) \notag \\
&\qquad + 8\alpha_3^{(3)}\left(\sqrt{\frac\pi2}\frac{U_{gg}}{\hbar\omega_g}\right)\left(\sqrt{\frac\pi2}\frac{U_{eg}}{\hbar\omega_{eg}}\left(\frac{\omega_g}{\omega_{eg}}\right)^{1/2}\right)^2
+ 2\alpha_3^{(3)}\left(\sqrt{\frac\pi2}\frac{U_{eg}}{\hbar\omega_{eg}}\left(\frac{\omega_g}{\omega_{eg}}\right)^{1/2}\right)^3 \notag \\
&\qquad - 2\alpha_{4,3}^{(3)}\left(\sqrt{\frac\pi2}\frac{U_{gg}}{\hbar\omega_g}\right)^2\left(\sqrt{\frac\pi2}\frac{U_{eg}}{\hbar\omega_{eg}}\left(\frac{\omega_g}{\omega_{eg}}\right)^{1/2}\right)
- 2\alpha_{4,3}^{(3)}\left(\sqrt{\frac\pi2}\frac{U_{gg}}{\hbar\omega_g}\right)\left(\sqrt{\frac\pi2}\frac{U_{eg}}{\hbar\omega_{eg}}\left(\frac{\omega_g}{\omega_{eg}}\right)^{1/2}\right)^2 \notag \\
&\qquad - 2\alpha_{4,3}^{(3)}\left(\sqrt{\frac\pi2}\frac{U_{eg}}{\hbar\omega_{eg}}\left(\frac{\omega_g}{\omega_{eg}}\right)^{1/2}\right)^3
- 4\alpha_5^{(3)}\left(\sqrt{\frac\pi2}\frac{U_{gg}}{\hbar\omega_g}\right)^2\left(\sqrt{\frac\pi2}\frac{U_{eg}}{\hbar\omega_{eg}}\left(\frac{\omega_g}{\omega_{eg}}\right)^{1/2}\right) \notag \\
&\qquad \left.- 10\alpha_5^{(3)}\left(\sqrt{\frac\pi2}\frac{U_{gg}}{\hbar\omega_g}\right)\left(\sqrt{\frac\pi2}\frac{U_{eg}}{\hbar\omega_{eg}}\left(\frac{\omega_g}{\omega_{eg}}\right)^{1/2}\right)^2
- 4\alpha_5^{(3)}\left(\sqrt{\frac\pi2}\frac{U_{eg}}{\hbar\omega_{eg}}\left(\frac{\omega_g}{\omega_{eg}}\right)^{1/2}\right)^3\right]\left(\frac{\omega_{egg}}{\omega_g}\right)^{3/2} + O(U^4) \label{eq:Uegg_Uexpansion}
\end{align}
and
\begin{align}
\frac{U_{eggg}}{\hbar\omega_{eggg}}
&= \left[24\alpha_{4,1}^{(3)}\left(\sqrt{\frac\pi2}\frac{U_{gg}}{\hbar\omega_g}\right)^2
\left(\sqrt{\frac\pi2}\frac{U_{eg}}{\hbar\omega_{eg}}
\left(\frac{\omega_g}{\omega_{eg}}\right)^{1/2}\right)
+ 24\alpha_{4,1}^{(3)}\left(\sqrt{\frac\pi2}\frac{U_{gg}}{\hbar\omega_g}\right)
\left(\sqrt{\frac\pi2}\frac{U_{eg}}{\hbar\omega_{eg}}
\left(\frac{\omega_g}{\omega_{eg}}\right)^{1/2}\right)^2\right. \notag \\
&\qquad + 30\alpha_{4,2}^{(3)}\left(\sqrt{\frac\pi2}\frac{U_{gg}}{\hbar\omega_g}\right)^2
\left(\sqrt{\frac\pi2}\frac{U_{eg}}{\hbar\omega_{eg}}
\left(\frac{\omega_g}{\omega_{eg}}\right)^{1/2}\right)
+ 12\alpha_{4,2}^{(3)}\left(\sqrt{\frac\pi2}\frac{U_{gg}}{\hbar\omega_g}\right)
\left(\sqrt{\frac\pi2}\frac{U_{eg}}{\hbar\omega_{eg}}
\left(\frac{\omega_g}{\omega_{eg}}\right)^{1/2}\right)^2 \notag\\
&\qquad + 6\alpha_{4,2}^{(3)}\left(\sqrt{\frac\pi2}\frac{U_{eg}}{\hbar\omega_{eg}}
\left(\frac{\omega_g}{\omega_{eg}}\right)^{1/2}\right)^3
- 42\alpha_5^{(3)}\left(\sqrt{\frac\pi2}\frac{U_{gg}}{\hbar\omega_g}\right)^2
\left(\sqrt{\frac\pi2}\frac{U_{eg}}{\hbar\omega_{eg}}
\left(\frac{\omega_g}{\omega_{eg}}\right)^{1/2}\right) \notag\\
&\qquad\left.- 24\alpha_5^{(3)}\left(\sqrt{\frac\pi2}\frac{U_{gg}}{\hbar\omega_g}\right)
\left(\sqrt{\frac\pi2}\frac{U_{eg}}{\hbar\omega_{eg}}
\left(\frac{\omega_g}{\omega_{eg}}\right)^{1/2}\right)^2
- 6\alpha_5^{(3)}\left(\sqrt{\frac\pi2}\frac{U_{eg}}{\hbar\omega_{eg}}
\left(\frac{\omega_g}{\omega_{eg}}\right)^{1/2}\right)^3\right]
\left(\frac{\omega_{eggg}}{\omega_g}\right)^{3/2}
+ O(U^4). \label{eq:Ueggg_Uexpansion}
\end{align}
\end{widetext}

\section{Sample preparation} \label{App:sample}
$^{174}$Yb BEC is produced by evaporative cooling in a crossed dipole trap operating at a wavelength of 532~nm.
Here, the atom number of BEC is about $3\times 10^4$.
The prepared atoms are finally introduced into a 3D cubic optical lattice that is operated at a wavelength of 532~nm through an optical superlattice configuration at an intermediate stage, as follows (see Fig.~\ref{Fig.5}):

\indent
(1) Atoms are loaded adiabatically into the optical lattice at a potential depth of $s_L = \left[(s_{\text{short}}^{(x)},\ s_{\text{long}}^{(x)}),\ s_{\text{short}}^{(y)},\ (s_{\text{short}}^{(z)},\ s_{\text{long}}^{(z)})\right]=\left[(0,\ 10),\ 25,\ (0,\ 10)\right]$ in 300~ms, where $s_{\text{short}}^{(x,y,z)}$ and $s_{\text{long}}^{(x,z)}$ are the potential depth of the short lattice (wavelength $\lambda_{\text{short}}=\ $532~nm) and the long lattice ($\lambda_{\text{long}}=\ $1064~nm) normalized by the recoil energy for each wavelength, respectively.

\indent
(2) We ramp up the short lattice to $s_L = \left[(5,\ 20),\ 25,\ (5,\ 20)\right]$ in 10~ms and trap atoms in an optical superlattice in a staggered phase (the relative phase between the short and long lattices is zero), where the potential difference between the two sites in the unit cell of the superlattice is maximized, for the $x$ and $z$ directions.

\indent
(3) The short lattice is quickly ramped up to the final potential depth of $s_L = \left[(25,\ 0),\ 25,\ (25,\ 0)\right]$ in 0.2~ms, while at the same time the long lattice is ramped down.
Here, we use the atom-number-projection technique \cite{PhysRevA.99.033609}, which is also helpful to increase the number of sites with large atomic occupancy.

After loading atoms into the optical lattice, the dipole trap is turned off to reduce spatial inhomogeneity of the trapping potential, and the atoms are trapped only by the optical lattice potential, followed by the irradiation of the excitation laser.\\
\indent
In the configuration of our optical lattice system, the trap angular frequencies for atoms in the ground ${^1}S_0$ and excited ${^3}P_2\ (m_J=+2)$ states are ${\bm \omega}_g = 2\pi\times(35.9,35.9,35.9)$~kHz and ${\bm \omega}_e = 2\pi\times(44.0,44.0,38.5)$~kHz, respectively, at the final potential depth of $25\,E_R$.
Note that the inhomogeneity of trap frequencies is estimated to be about 0.2~\%.

\section{Details of the data analysis of spectroscopic results} \label{App:Analysis}
First, we determine the resonance position and its fitting error for every atomic occupancy $n$ in each scan indexed as $i$ by fitting the spectrum with a multi-Gaussian function given by
\begin{equation}
  \sum_n a_{n,i}\exp\left[ -\frac{(x-x_{n,i})^2}{2\sigma_n^2}\right]+b_i, \label{fitfunc}
\end{equation}
where $x$ represents the laser frequency, $x_{n,i}$ the resonance position of the $n$ occupancy, $\sigma_n$ the width of the Gaussian function taken to be common to each scan, $b_i$  the offset, and $a_{n,i}$ the amplitude of the Gaussian function.
We perform this procedure for several, typically six, scans taken at the same magnetic field.
Note that the quantity of interest in these measurements is not the absolute resonance position but the relative shift between the resonances.
Thus, in the following, we consider $x_{\text{shift},n,i}$, the shift of the resonance frequency for $n$ occupancy with respect to $n=1$, given by $x_{\text{shift},n,i}=x_{n,i}-x_{n=1,i}$.
Then, as a result of the fittings, we obtain $x_{\text{shift},n}$, a weighted average of the resonance position shifts, given as $x_{\text{shift},n} = \sum_i(w_{n,i}\,x_{\text{shift},n,i})/\sum_i w_{n,i}$, using the weight $w_{n,i}=1/\sigma_{\text{shift},n,i}^2$ where $\sigma_{\text{shift},n,i}$ is the fitting error.

We also obtain $\sigma_{n, \text{StdDev}}$, a weighted standard deviation of the resonance position shifts, given as $\sigma_{n, \text{StdDev}} = \sqrt{\sum_i[w_{n,i}\,(x_{\text{shift},n,i}-x_{\text{shift},n})^2]/\sum_i w_{n,i}}$, to represent the error of the resonance position shifts for every $n$ at a certain magnetic field
\footnote{
The use of $\sigma_{n, \text{StdDev}}$ as a representative for the error is justified if $\sigma_{n, \text{StdDev}}$ is not much smaller than $\sigma_{n,\text{ErrProp}}=1/\sqrt{\sum_i w_{n,i}}$ calculated from the fitting errors $\sigma_{\text{shift}, n, i}$.
However, this is not the case for the resonances of $n=4$ and $n=5$ at $B_0=232$~mG, $n=3$ at $280$~mG, and $n=5$ at $327$~mG, where we use $\sigma_{n, \text{ErrProp}}$ instead of $\sigma_{n, \text{StdDev}}$ to represent the error.
}.
In addition, the spectrum suffers from long-term laser frequency drifts, which also affect the relative frequency shifts.
This effect is compensated for by a linear interpolation with the reference to the resonance position of $n=2$ or $n=1$.
Here, we consider the difference between the resonance positions of $n=1$ or $n=2$ and the linear fit to them, as the error $\sigma_{\text{Drift}}$ of the drift compensation.
Namely, $\sigma_{\text{Drift}} = \sqrt{\sum_i[w'_{i}\,(f_{\text{diff},i}-f_{\text{avg}})^2]/\sum_i w'_{i}}$, where $f_{\text{diff},i}$ is the frequency difference between the resonance positions of $n=1$ or $n=2$ and the linear fit to them in each scan indexed as $i$, $f_{\text{avg}}$ is the average of $f_{\text{diff},i}$ weighted by the fitting errors $\sigma'_{\text{shift},n,i}$ of the resonance positions of $n=1$ or $n=2$, and $w'_{i}=1/\sigma'^{2}_{\text{shift},n,i}$.
Note that $\sigma_{\text{Drift}}$ is typically about 1.0--1.5~kHz.
In the end, the value calculated with the following formula is used as the total error $\sigma_{\text{total}}$:
\begin{equation}
  \sigma_{\text{total}} \coloneqq \sqrt{\sigma_{n, \text{StdDev}}^2+\sigma_{\text{Drift}}^2}.
\end{equation}
\indent
We note that some of the $n=1$ resonance spectra show a tail (see Fig.~\ref{Fig.1} (upper)), which is not represented by a simple Gaussian function.
This additional structure is explained as the excitation of the center-of-mass motion for the $n=3$ occupancy site, or $n=2$, depending on the magnetic field strength.
We estimate the associated error in the determination of the $n=1$ resonance to be typically less than 0.3~kHz and not greater than 0.5~kHz.
Since this is about two-orders-of magnitude smaller than the typical frequency shifts, we neglect this effect in the analysis.

\section{Scattering length dependence of two-body interaction energy} \label{App:Scattering}

\begin{figure*}[t]
\includegraphics[width=0.49\textwidth]{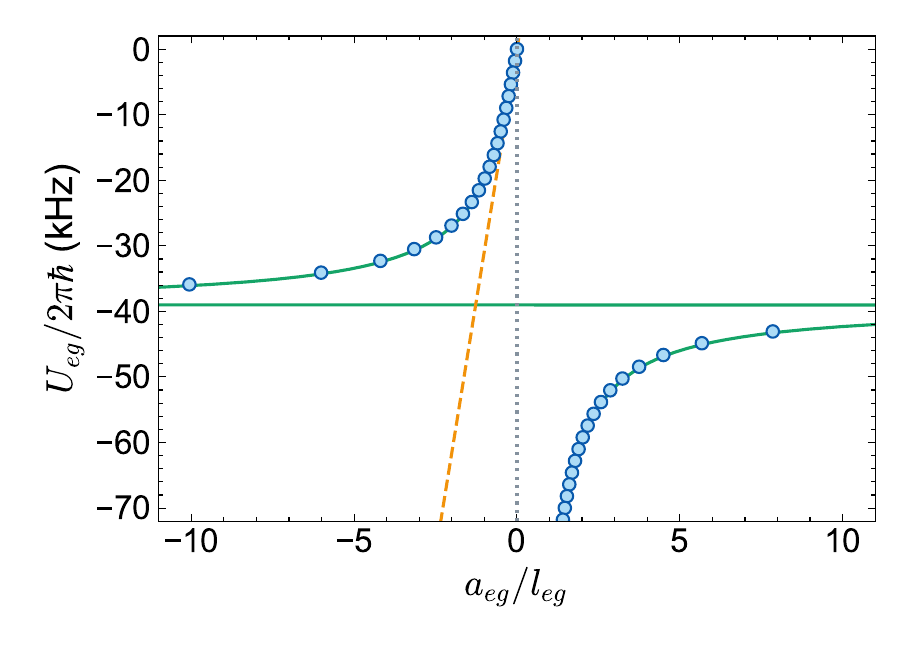}\hfill
\includegraphics[width=0.49\textwidth]{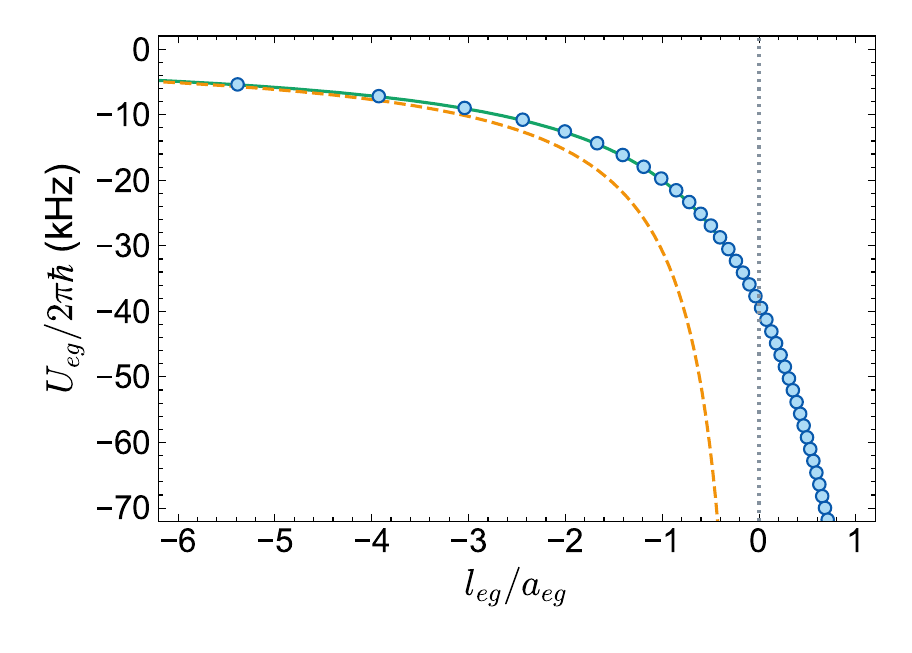}
\caption{\label{fig:Ueg}
$U_{eg}/(2\pi\hbar)$ in units of kHz as functions of $a_{eg}/l_{eg}$ (left) and $l_{eg}/a_{eg}$ (right) with $l_{eg}=\sqrt{\hbar/(m\omega_{eg})}=38.6$~nm.
The blue points are numerical results from Eq.~(\ref{eq:integral}), the orange dashed curve is from the first-order perturbation theory in Eq.~(\ref{eq:perturbation}), and the green solid curve assumes Eq.~(\ref{eq:approximation}) under the equal frequency approximation.
The resonance ($l_g/a_{eg}=0$) is located at $U_{eg}/(2\pi\hbar)=-38.8$ kHz from the numerical results and at $U_{eg}/(2\pi\hbar)=-39.0$ kHz from the equal frequency approximation.}
\end{figure*}

Two $^{174}$Yb atoms in the ground ${^1}S_0$ and excited ${^3}P_2$ states trapped by unequal and anisotropic harmonic potentials are described by the Schr\"odinger equation,
\begin{align}
& \Biggl[\underbrace{\sum_{\sigma=g,e}\sum_{i=x,y,z}
\left(-\frac1{2m}\frac{\d^2}{\d r_{\sigma i}^2}
+ \frac{m}{2}\omega_{\sigma i}^2r_{\sigma i}^2\right)}_{\hat{H}_0} \notag\\
& + \frac{4\pi a_{eg}}{m}\delta_\reg(\r_g-\r_e)\Biggr]
\psi(\r_g,\r_e) = E_{eg}\psi(\r_g,\r_e),
\end{align}
where $m=2.89\times10^{-25}$ kg is the mass of $^{174}$Yb atoms, $\bm\omega_{g(e)}$ are the trap angular frequencies for the ground (excited) state, $a_{eg}$ is the scattering length between them, and $\delta_\reg(\r)\coloneqq\delta(\r)\frac\d{\d|\r|}|\r|$ with $\hbar=1$ throughout this paragraph.
The corresponding Lippmann-Schwinger equation below the threshold at $E_{eg}<E_g+E_e=\sum_{i=x,y,z}\frac{\omega_{gi}+\omega_{ei}}{2}$ reads
\begin{align}
\psi(\r_g,\r_e) = \frac{4\pi a_{eg}}{m}\int\!d\r'G(\r_g,\r_e;\r',\r')\chi(\r'),
\end{align}
where
\begin{align}
G(\r_g,\r_e;\r_g',\r_e') &\coloneqq \<\r_g,\r_e|\frac1{E_{eg}-\hat{H}_0}|\r_g',\r_e'\>
\end{align}
is the Green function and
\begin{align}
\chi(\r) \coloneqq \lim_{\r_g,\r_e\to\r}\frac\d{\d|\r_g-\r_e|}[|\r_g-\r_e|\psi(\r_g,\r_e)]
\end{align}
is the regularized wave function at a coincident point.
Because of
\begin{align}
\lim_{\r_g,\r_e\to\r}\psi(\r_g,\r_e)
= \left[1-\frac{a_{eg}}{|\r_g-\r_e|}\right]\chi(\r) + O(|\r_g-\r_e|),
\end{align}
we obtain
\begin{align}\label{eq:integral}
\chi(\r) = \frac{4\pi a_{eg}}{m}\int\!d\r'G_\reg(\r;\r')\chi(\r')
\end{align}
with
\begin{align}
& G_\reg(\r;\r') \notag\\
&\coloneqq \lim_{\r_g,\r_e\to\r}
\left[G(\r_g,\r_e;\r',\r') + \frac{m}{4\pi|\r_g-\r_e|}\delta(\r-\r')\right]
\end{align}
being the regularized Green function.

The Green function for $U_{eg}=E_{eg}-E_g-E_e<0$ is provided by
\begin{align}
& G(\r_g,\r_e;\r_g',\r_e')
= -\int_0^\infty\!d\tau\,\<\r_g,\r_e|e^{(E_{eg}-\hat{H}_0)\tau}|\r_g',\r_e'\> \\
&= -\int_0^\infty\!d\tau\,e^{U_{eg}\tau} \notag\\
&\quad \times \prod_{i=x,y,z}
\Biggl[\sqrt{\frac{m\omega_{gi}e^{\omega_{gi}\tau}}{2\pi\sinh\omega_{gi}\tau}}\,
e^{-\frac{m\omega_{gi}}{2}\frac{(r_{gi}^2+r_{gi}'^2)\cosh\omega_{gi}\tau
- 2r_{gi}r_{gi}'}{\sinh\omega_{gi}\tau}} \notag\\
&\qquad \times \sqrt{\frac{m\omega_{ei}e^{\omega_{ei}\tau}}{2\pi\sinh\omega_{ei}\tau}}\,
e^{-\frac{m\omega_{ei}}{2}\frac{(r_{ei}^2+r_{ei}'^2)\cosh\omega_{ei}\tau
- 2r_{ei}r_{ei}'}{\sinh\omega_{ei}\tau}}\Biggr].
\end{align}
Because of
\begin{align}
\int_0^\infty\!d\tau\left(\frac{m}{4\pi\tau}\right)^{3/2}
e^{-\frac{m(\r_g-\r_e)^2}{4\tau}} = \frac{m}{4\pi|\r_g-\r_e|},
\end{align}
we obtain
\begin{align}
& G_\reg(\r;\r') \notag\\
&= -\int_0^\infty\!d\tau\Biggl[e^{U_{eg}\tau}\prod_{i=x,y,z}
\sqrt{\frac{m\omega_{gi}e^{\omega_{gi}\tau}}{2\pi\sinh\omega_{gi}\tau}
\frac{m\omega_{ei}e^{\omega_{ei}\tau}}{2\pi\sinh\omega_{ei}\tau}} \notag\\
&\quad \times e^{-\frac{m\omega_{gi}}{2}
\frac{(r_i^2+r_i'^2)\cosh\omega_{gi}\tau-2r_ir_i'}{\sinh\omega_{gi}\tau}
- \frac{m\omega_{ei}}{2}
\frac{(r_i^2+r_i'^2)\cosh\omega_{ei}\tau-2r_ir_i'}{\sinh\omega_{ei}\tau}} \notag\\
&\qquad - \left(\frac{m}{4\pi\tau}\right)^{3/2}\delta(\r-\r')\Biggr]
\end{align}
for the regularized Green function.
With this expression, the integral equation in Eq.~(\ref{eq:integral}) can be solved numerically by expanding $\chi(\r)$ in terms of orthogonal polynomials such as Hermite polynomials.
The resulting two-body interaction energy $U_{eg}$ as a function of $a_{eg}$ is shown in Fig.~\ref{fig:Ueg} for trap frequencies of
\begin{align}
\bm\omega_g &= 2\pi \times (35.9,35.9,35.9)\ \mathrm{kHz}, \\
\bm\omega_e &= 2\pi \times (44.0,44.0,38.5)\ \mathrm{kHz},
\end{align}
which are determined as the average band gaps from the optical lattices with $\lambda=532$~nm, $\s_g=25\times(1,1,1)$, and $\s_e=25\times(1.43,1.43,1.132)$.
Figure~\ref{fig:Ueg} also shows the result from the first-order perturbation theory predicting
\begin{align}\label{eq:perturbation}
U_{eg} = \frac{4\pi a_{eg}}{m}\prod_{i=x,y,z}
\sqrt{\frac{m}{\pi}\frac{\omega_{gi}\omega_{ei}}{\omega_{gi}+\omega_{ei}}} + O(a_{eg}^2).
\end{align}

In the special case of equal and isotropic trap frequency $\omega_{eg}$ for two atoms, Busch {\it et al.} in Ref.~\cite{Busch1998} finds
\begin{align}\label{eq:approximation}
\frac1{a_{eg}}\sqrt{\frac\hbar{m\omega_{eg}}}
= \frac{\sqrt2\,\Gamma\!\left(-\frac{U_{eg}}{2\hbar\omega_{eg}}\right)}
{\Gamma\!\left(-\frac12-\frac{U_{eg}}{2\hbar\omega_{eg}}\right)},
\end{align}
whereas anisotropic cases have been studied in Refs.~\cite{PhysRevA.71.050701,PhysRevA.74.022712,Liang_2008,PhysRevA.101.053624}.
The two-body interaction energy $U_{eg}$ assuming an average trap frequency of
\begin{align}
\omega_{eg} = \frac13\sum_{i=x,y,z}\frac{\omega_{gi}+\omega_{ei}}{2}
= 2\pi \times 39.0\ \mathrm{kHz}
\end{align}
is shown in Fig.~\ref{fig:Ueg} as a function of $a_{eg}$.
The resulting $U_{eg}$ under such equal frequency approximation is found to be in close agreement with the numerical results, which fully take into account the unequal and anisotropic harmonic potentials for two atoms as in the experimental situation of this study.
Because the relative difference is less than 1\,\%, the equal frequency approximation is employed to plot the theoretical curve in Fig.~\ref{Fig.2}(b) of the main text as well as to compute effective multi-body interactions in Sec.~\ref{Effective multi-body interactions with perturbation theory}.
We also note that the contribution of the effective range compared with the dominant term in the $s$-wave scattering amplitude is estimated to be smaller than 1/6 for the scattering length we are working on \cite{RevModPhys.82.1225,PhysRevA.77.012719}, so that the effective range is neglected in our analysis.

\begin{figure}[t]
\includegraphics[width=8.6cm]{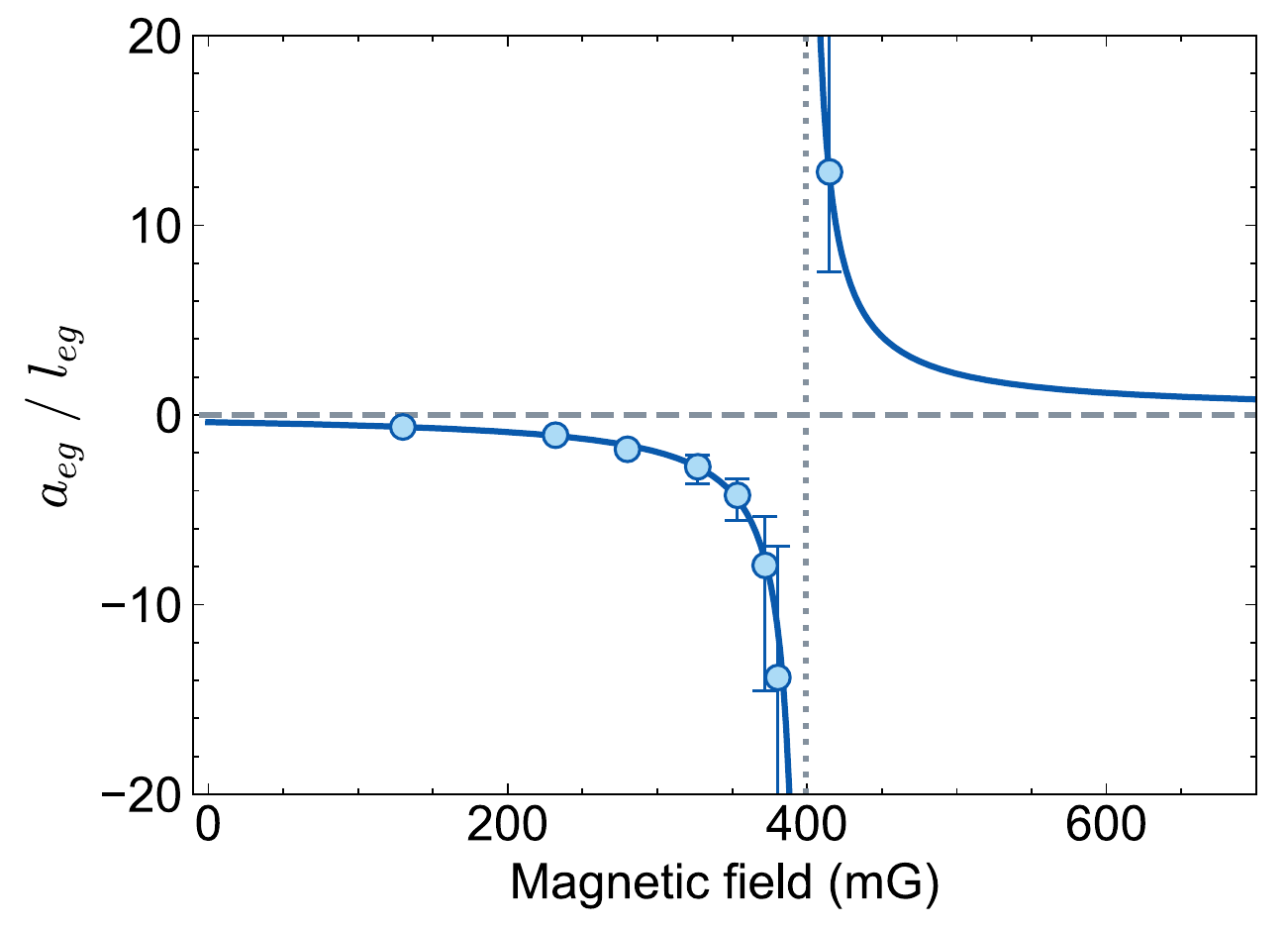}
\caption{\label{fig:a_egvsB}
Magnetic field variation of the scattering length $a_{eg}$ in the Feshbach resonance.
The solid curve denotes a fit on all plots with a fitting function $a_{eg}(B) = a_{bg} - \Delta'/(B-B_0)$, with $a_{bg}=0.13(0.16)\,l_{eg}$, $\Delta'=-206(45)$~mG, and $B_0=398(13)$~mG.
The vertical error bars at each magnetic field strength are the same as the horizontal error bars in Fig.~\ref{Fig.2}(b) of the main text.}
\end{figure}

On the other hand, the interaction energy of two atoms both in the ground state reads
\begin{align}
U_{gg} = 2\pi\hbar \times 4.04\ \mathrm{kHz}
\end{align}
for $\omega_g=2\pi\times35.9$ kHz and $a_{gg}=5.55$~nm~\cite{PhysRevA.77.012719}.
Now that we know the two-body interaction energies $U_{gg}$ and $U_{eg}$ as a function of $a_{eg}$, we can determine the scattering length $a_{eg}$ in the Feshbach resonance from the experimental data for $\nu_2-\nu_1=(U_{eg}-U_{gg})/(2\pi\hbar)$ in Fig.~\ref{Fig.2}(a) of the main text.
The resulting magnetic field variation of $a_{eg}$ is shown in Fig.~\ref{fig:a_egvsB}
\footnote{The magnetic field strength at the Feshbach resonance point is obtained as $B_0=398(13)$~mG in this study, whereas that is $360(10)$~mG in the previous study \cite{PhysRevLett.110.173201}.
We do not currently have a clear explanation for this discrepancy.
Because the magnetic field is properly calibrated as mentioned in the main text, we describe the Feshbach resonance based on the present calibration of the magnetic field in this study.}.\\

\section{Derivation of linear relationships between frequency shifts} \label{App:Derivation}
The effective zero-dimensional Hamiltonian up to five atoms including up to one excited-state atom is given by Eq.~(\ref{eq:effective}).
The energies of $|g\>=\hat{b}_g^\+|0\>$, $|gg\>=(\hat{b}_g^{\+2}/\sqrt{2!})|0\>$, $|ggg\>=(\hat{b}_g^{\+3}/\sqrt{3!})|0\>$, $|gggg\>=(\hat{b}_g^{\+4}/\sqrt{4!})|0\>$, $|ggggg\>=(\hat{b}_g^{\+5}/\sqrt{5!})|0\>$, $|e\>=\hat{b}_e^\+|0\>$, $|eg\>=\hat{b}_e^\+\hat{b}_g^\+|0\>$, $|egg\>=(\hat{b}_e^\+\hat{b}_g^{\+2}/\sqrt{2!})|0\>$, $|eggg\>=(\hat{b}_e^\+\hat{b}_g^{\+3}/\sqrt{3!})|0\>$, $|egggg\>=(\hat{b}_e^\+\hat{b}_g^{\+4}/\sqrt{4!})|0\>$ states are provided by
\begin{align}
E_g &= U_g, \label{eq:Eg} \\
E_{gg} &= 2U_g+U_{gg}, \\
E_{ggg} &= 3U_g+3U_{gg}+U_{ggg}, \\
E_{gggg} &= 4U_g+6U_{gg}+4U_{ggg}+U_{gggg}, \\
E_{ggggg} &= 5U_g+10U_{gg}+10U_{ggg}+5U_{gggg}+U_{ggggg}, \\
E_e &= U_e, \\
E_{eg} &= U_e+U_g+U_{eg}, \\
E_{egg} &= U_e+2U_g+2U_{eg}+U_{gg}+U_{egg}, \\
E_{eggg} &= U_e+3U_g+3U_{eg}+3U_{gg}+3U_{egg}+U_{ggg}\notag\\
&\quad +U_{eggg}, \\
E_{egggg} &= U_e+4U_g+4U_{eg}+6U_{gg}+6U_{egg}+4U_{ggg}\notag\\
&\quad +4U_{eggg}+U_{gggg}+U_{egggg}. \label{eq:Eegggg}
\end{align}
Therefore, the transition energies are found to be
\begin{align}
h\nu_1 &= E_e-E_g = U_e-U_g, \\
h\nu_2 &= E_{eg}-E_{gg} = U_e-U_g+U_{eg}-U_{gg}, \\
h\nu_3 &= E_{egg}-E_{ggg}\notag\\
&= U_e-U_g+2U_{eg}-2U_{gg}+U_{egg}-U_{ggg}, \\
h\nu_4 &= E_{eggg}-E_{gggg}\notag\\
&= U_e-U_g+3U_{eg}-3U_{gg}+3U_{egg}-3U_{ggg}\notag\\
&\quad +U_{eggg}-U_{gggg}, \\
h\nu_5 &= E_{egggg}-E_{ggggg}\notag\\
&= U_e-U_g+4U_{eg}-4U_{gg}+6U_{egg}-6U_{ggg}\notag\\
&\quad +4U_{eggg}-4U_{gggg}+U_{egggg}-U_{ggggg},
\end{align}
so that
\begin{align}
h\nu_2-h\nu_1 &= U_{eg}-U_{gg}, \\
h\nu_3-h\nu_1 
&= 2(h\nu_2-h\nu_1)+U_{egg}-U_{ggg}, \label{eq:nu3-nu1}\\
h\nu_4-h\nu_1 
&= 3(h\nu_3-h\nu_2)+U_{eggg}-U_{gggg}, \label{eq:nu4-nu1}\\
h\nu_5-h\nu_1
&= 6(h\nu_3-h\nu_1)-8(h\nu_2-h\nu_1)\notag\\
&\quad +4U_{eggg}-4U_{gggg}+U_{egggg}-U_{ggggg} \label{eq:nu5-nu1_4-5body}\\
&= 4(h\nu_4-h\nu_1)-6(h\nu_3-h\nu_1)+4(h\nu_2-h\nu_1)\notag\\
&\quad +U_{egggg}-U_{ggggg}.\label{eq:nu5-nu1_5body}
\end{align}
If the three-body and four-body interactions are neglected in Eqs.~(\ref{eq:nu3-nu1}) and (\ref{eq:nu4-nu1}), we obtain Eqs.~(\ref{eq:linear3}) and (\ref{eq:linear4}) of the main text, respectively.
Also, if the interactions with four and five bodies or more are neglected in Eqs.~(\ref{eq:nu5-nu1_4-5body}) and (\ref{eq:nu5-nu1_5body}), we obtain Eqs.~(\ref{eq:linear5'}) and (\ref{eq:linear5''}) of the main text, respectively.

\section{Revisiting the results in previous studies} \label{App:Revisiting}
\subsection{Chi-square analysis of the data in S.~Will {\it et al.} (2010) \cite{Will2010}}
In this previous study of performing time-resolved observations of quantum phase collapse and revival, the measured effective four-body interaction strength $U_4$ at each optical lattice potential depth is shown in Fig.~3b of that paper.
Here, we read the values of $U_4$ and its errors from the figure and perform a chi-square test with the $\chi^2$ function defined below, assuming that $U_4$ takes zero:
\begin{align}
  \chi^2 \coloneqq \sum_i \left(\frac{\bar{y}_i}{\sigma_{\bar{y}_i}}\right)^2,
\end{align}
where $i$ is the data index and $(\bar{y}_i, \sigma_{\bar{y}_i})$ are the measured quantities and their uncertainties for $U_4$.
In this $\chi^2$ function, we obtain $\chi^2=1.72$ ($\text{dof}=5$) and $p\text{-value}=0.886$, which corresponds to a significance of $0.14\,\sigma$.
This indicates that $U_4$ is consistent with zero.

\subsection{G.~K.~Campbell {\it et al.} (2006) \cite{Campbell2006}}
In this previous study of imaging Mott insulators in a three-dimensional optical lattice with microwave spectroscopy, the measured dependence of the resonance position on the atomic occupation number $n$ is shown in Fig.~2B (circle symbols) of Ref.~\cite{Campbell2006} up to $n=5$.
The resonance positions corresponding to each $n$ are read from the figure (no readings are available for error bars) and checked to see if the relationships of Eqs.~(\ref{eq:linear4}) and (\ref{eq:linear5'}), valid in the case of no effective four- and five-body interactions, are satisfied.
As a result, the left- and right-hand sides of Eq.~(\ref{eq:linear4}) are $-88$~Hz and $-87$~Hz, respectively, and the left- and right-hand sides of Eq.~(\ref{eq:linear5'}) are $-111$~Hz and $-112$~Hz, respectively.
This means that both $n=4$ and $n=5$ shifts in this previous study can be explained by considering only up to effective three-body interactions.

\subsection{A.~Goban {\it et al.} (2018) \cite{Goban2018}}
In this previous study of performing high-resolution clock spectroscopy for fermionic $^{87}$Sr atoms in a three-dimensional optical lattice, the measured dependence of the resonance position up to the atomic occupation number $n=5$ is shown in Fig.~3 (red circles) of Ref.~\cite{Goban2018}.
Here, the authors find that the measured values deviate from the calculated values when only two-body interactions are considered, while they are consistent with the perturbative calculation results that consider up to effective three-body interactions.
They also mention in the supplementary information that the contribution of effective four-body interactions in their system is so small that they can be ignored, and in Fig.~S1 the measured resonance positions up to $n=5$ in the wide range of mean trap depths show `good agreement' with the calculated values considering up to the effective three-body interactions.

\subsection{L.~Franchi {\it et al.} (2017) \cite{Franchi_2017}}
In this previous study of measuring the scattering properties of ultracold $^{174}$Yb bosons with high-resolution spectroscopy in a three-dimensional optical lattice, the measured dependence of the resonance positions for doubly ($n=2$, red circles) and triply ($n=3$, blue circles) occupied sites on the potential depth of the 3D lattice is shown in Fig.~3 of that paper.
Here, only up to the effective three-body interactions appear at most, and the authors find that the observed resonance positions show `very good agreement' with the fit including the effective three-body interactions calculated by the perturbation theory.


\providecommand{\noopsort}[1]{}\providecommand{\singleletter}[1]{#1}%

\end{document}